\documentclass{article}[11pt]
\usepackage{amssymb}
\newtheorem{definition}{Definition}
\newtheorem{lemma}{Lemma}[section]

\newtheorem{corollary}[lemma]{Corollary}
\newtheorem{theorem}[lemma]{Theorem}
\newtheorem{proposition}[lemma]{Proposition}
\newcommand{\shuffle}{\,\hbox{\mbox{$\scriptstyle\sqcup{\mskip-4mu}\sqcup$}}\,}
\newcommand{\Q}{\mathbb{Q}}

\newtheorem{exampleA}{Example}
\newtheorem{remarkA}{Remark}

\newenvironment{proof}[1][Proof.]
    {\addvspace{\bigskipamount}\noindent\emph{#1} }
    {\par\addvspace{\bigskipamount}}

\newcommand{\qed}{\ifmmode\raisebox{3.3pt}{\fbox{}}\else{\ifhmode
\unskip\fi\nobreak\hfil\penalty50\hskip1em\null\nobreak\hfil\raisebox
{3.3pt}{\fbox{}}\parfillskip=0pt\finalhyphendemerits=0\endgraf}\fi}

\newcommand{\IM}{\mathop{\mathrm{Im}}\nolimits}

\newenvironment{proofWithQed}{\par\noindent {\it Proof.} \rm}{\ ~~~\qed}
\newenvironment{proof2}[1]{\par\noindent {\it Proof of #1.} \rm}{\ ~~~\qed}

\newcommand{\LL}[1]{{\rm L}_{\vdash_{\{ #1\}}}^{\epsilon}}
\newcommand{\LLSet}[1]{{\rm L}_{\vdash_{ #1}}^{\epsilon}}
\newcommand{\qo}[1]{\vdash_{\{#1\}}}
\newcommand{\qoEtoile}[1]{\vdash_{\{#1\}}^*}
\newcommand{\qoEtoileSet}[1]{\vdash_{#1}^*}
\newcommand{\mod}{{\rm ~mod~}}

\begin{document}

\thispagestyle{empty}
\begin{center}
LaRIA~: Laboratoire de Recherche en Informatique d'Amiens\\
Université de Picardie Jules Verne -- CNRS FRE 2733\\
33, rue Saint Leu, 80039 Amiens cedex 01, France\\
Tel : (+33)[0]3 22 82 88 77\\
Fax : (+33)[0]03 22 82 54 12\\
\underline{http://www.laria.u-picardie.fr}
\end{center}

\vspace{7cm}

\begin{center}
\parbox[t][5.9cm][t]{10cm}
{\center

{\bf Well quasi-orders and\\
the shuffle closure of finite sets

\medskip

F. D'Alessandro$^{\rm a}$, G. Richomme$^{\rm b}$, S. Varrichio$^{\rm c}$
}
\bigskip

\textbf{L}aRIA \textbf{R}ESEARCH \textbf{R}EPORT~: LRR 2006-06\\
(July 2006)
}
\end{center}

\vfill

\hrule depth 1pt \relax

\medskip

\noindent
$^a$ Universit\`a di Roma ``La Sapienza'', dalessan@mat.uniroma1.it\\
$^b$ LaRIA, Université de Picardie Jules Verne, gwenael.richomme@u-picardie.fr\\
$^c$ Universit\`a di Roma ``Tor Vergata'', varrich@mat.uniroma2.it

\vspace{-2cm}
\pagebreak

 \pagestyle{plain}

\title{Well quasi-orders and the shuffle closure of finite sets
\thanks{This work was partially supported by MIUR project { \tt
``Linguaggi formali e automi: teoria e applicazioni''.}}  \thanks{An
extended abstract of this paper was presented at the conference
DLT'2006, Santa Barbara, June 2006, Lecture Notes in Computer Science
4036 (2006), pages 260-269, Springer-Verlag, Berlin.}
}

\author{Flavio D'Alessandro \\
        Dipartimento di Matematica, \\
Universit\`a di Roma ``La Sapienza'' \\
  Piazzale Aldo Moro 2, 00185 Roma,  Italy. \\
  e-mail: dalessan@mat.uniroma1.it,\\
\texttt{http://www.mat.uniroma1.it/people/dalessandro}
\and Gw\'ena\"el Richomme \\
   LaRIA, UPJV \\
  33 Saint Leu, 80039  Amiens Cedex 01, France. \\
  e-mail: gwenael.richomme@u-picardie.fr, \\
   \texttt{http://www.laria.u-picardie.fr/{\small $\sim$}richomme/ }
 \and Stefano Varricchio \\
Dipartimento di Matematica, \\
 Universit\`a  di Roma ``Tor Vergata'', \\
  via della Ricerca Scientifica, 00133 Roma, Italy. \\
  e-mail: varricch@mat.uniroma2.it, \\
\texttt{http://www.mat.uniroma2.it/{\small $\sim$}varricch/ }}

\date{\today}
 
\maketitle

\begin{abstract}
    Given a set $I$ of word, the set $L^{\epsilon}_{\vdash_I}$ of all
    words obtained by the shuffle of (copies of) words of $I$ is
    naturally provided with a partial order: for $u, v$ in
    $L^{\epsilon}_{\vdash_I}$, $u \vdash^*_I v$ if and only if $v$ is
    the shuffle of $u$ and another word of
    $L^{\epsilon}_{\vdash_I}$. In \cite{FS05}, the authors have opened
    the problem of the characterization of the finite sets $I$ such
    that $\vdash_I^*$ is a well quasi-order on
    $L^{\epsilon}_{\vdash_I}$. In this paper we give an answer in the
    case when $I$ consists of a single word $w$.
\end{abstract}

\noindent
\textbf{Keywords:} formal languages, well quasi-orders, shuffle

\section{Introduction}

A  {\em quasi-order}  on a set  $S$  is called a  {\em well
quasi-order} ({\em wqo}) if every non-empty subset   $X$  of $S$ has
at least one minimal element in $X$ but no more than a finite number
of (non-equivalent) minimal elements. Well quasi-orders have been
widely investigated in the past. We recall the celebrated Higman
and Kruskal results \cite{hig,kru}.
 Higman gives a very general theorem on division orders  in abstract algebras
 from which one derives that the {\em subsequence ordering} in free monoids
 is a wqo. Kruskal extends Higman's result, proving that certain embeddings on finite trees
 are well quasi-orders. Some remarkable  extensions of the Kruskal theorem are given in  \cite{int,pue}.

In the last years many papers have been devoted to the application
of wqo's to formal language theory
\cite{DV2003,DV2004,DELVAR2,LIBRODV,ito,jan,roz,har,sal}.

Recently, in the theory of language equations, remarkable
 results based on wqo's have been obtained by M. Kunc 
\cite{KUNC1}. These results have
been culminating in the negative solution of the famous
conjecture by Conway stating the regularity of the maximal solutions
of the {\em commutative language equation} $XL = LX$ where $L$ is a finite language
of words \cite{KUNC2}.  

In \cite{roz}, a remarkable class of grammars, called {\em unitary
grammars}, has been introduced in order to study the relationships
between the classes of context-free  and regular languages. If $I$
is a finite set of words then we can consider the set of productions
$$\{\epsilon \rightarrow u \mid u\in I\}$$ and
the  derivation relation
  $\Rightarrow^*_I$ of the  semi-Thue system associated with
$I$. The language generated by the unitary grammar
associated with $I$ is ${\rm L}_{{I}} ^{\epsilon} = \{ w \in A^* \mid \epsilon
\Rightarrow^*_I w\}$. Unavoidable sets of words are characterized in
terms of  the wqo property of  the unitary grammars. Precisely it is
proved that $I$ is unavoidable if and only if the  derivation
relation $\Rightarrow^*_I$ is a wqo.

    In
\cite{hau}, Haussler investigated the relation
 $\vdash_I^*$ defined as the
 transitive and reflexive closure of $\vdash_I$
    where, for every pair $w, v$ of words,   $v \vdash_I w  $ if
 $$v = v_1v_2 \cdots v_{n+1},$$
 $$w =  v_1a_1v_2a_2 \cdots v_na_nv_{n+1},$$
where the  $a_i$'s are letters, and
 $a_1a_2 \cdots a_n \in I$.
In particular, a characterization of the wqo property of
$\vdash_I^*$  in terms of subsequence unavoidable sets of words was
given in \cite{hau}.
Let  $\LLSet{I}$ be  the set of all words derived from
the empty word by applying $\vdash^*_I$.

A remarkable result  proved in \cite{DV2004} states that for any
finite set $I$  the derivation relation $\vdash_I^*$ is a wqo on the
language ${\rm L}_{{I}} ^{\epsilon}$.
It is also proved that, in general, $\Rightarrow^*_I  $ is
not a wqo on ${\rm L}_{{I}} ^{\epsilon}$ and $\vdash_I^*$ is not a wqo on
$\LLSet{I}$. In \cite{FS05}  the authors characterize
the finite sets $I$ such that $\Rightarrow^*_I$ is a wqo on
${\rm L}_{{I}} ^{\epsilon}$. Moreover, they have left the following problem open:
{\em characterize the finite sets $I$ such that $\vdash_I^*$ is   a wqo
on $\LLSet{I}$.} In this paper we give an answer in the
case when $I$ consists of a single word $w$.

In this context, it is worth noticing that in \cite{FS05}
 the authors prove that $\vdash ^* _{\{w\}}$ is not a wqo on
$\LL{w}$ if $w = abc$.  
A simple argument allows one to extend the result above in the case that
$w = a^ib^jc^h$, $i,j,h\geq 1$. 
By using Lemma \ref{fs0}, this implies that
if a word $w$ contains three distinct letters at least, then
$\vdash ^* _{\{w\}}$ is not a wqo on
$\LL{w}$.  
Therefore, in order to prove our main result, we can focus
our attention to the case where $w$ is a word on the binary 
alphabet $\{a, b\}$.
Let
$E$ be the  exchange morphism ($E(a) = b$, $E(b) = a$),  and let
$\tilde{w}$ be the mirror image of $w$.

\begin{definition}\label{badw}
A word $w$ is called \emph{bad} if one of the words $w$, $\tilde{w}$, $E(w)$ and
$E(\tilde{w})$
has a factor of one of the two following forms\\
\begin{eqnarray}
\label{form1}
& a^kb^h & with ~ k, h \geq 2\\
\label{form2} & a^kba^lb^m &  with ~ k > l \geq 1, m \geq 1
\end{eqnarray}
 A word $w$ is called \emph{good} if it is not bad.
\end{definition}
Although it is immediate that a word $w$ is bad if and only if 
one of the words $w$, $\tilde{w}$, $E(w)$ and
$E(\tilde{w})$ contains a factor of the form $a^2b^2$¥ or
$a^{{k+1}}ba^{k}b$, with $k\geq 1$ it will be useful to consider
the definition as above. ¥
Morever we observe that, by Lemma \ref{goodWords} a word is good if and
only if it is a factor of $(ba^n)^{\omega}$  or $(ab^n)^{\omega}$
for some $n \geq 0$. The main result of our paper is the following.

\begin{theorem}
\label{mainTheorem} Let $w$ be a word over the alphabet 
$\{a, b\}$. The derivation relation
$\qoEtoile{w}$ is a wqo  on  $\LL{w}$ if and only if  $w$ is
good.
\end{theorem}

We assume the reader to be familiar with the 
basic theory of combinatorics on words  as well as
with the theory of well quasi-orders
({\em see also } \cite{LIBRODV, LIBROLOT}).
Now let us recall the following theorem which gives a useful characterization of
the concept of well quasi-order.
\begin{theorem}\label{4TLethat}
 Let   $S$  be a set  quasi-ordered  by $\leq$.
 The following conditions are equivalent:
\begin{enumerate}
\item[i.]  $\leq$   is a well quasi-order;
\item[ii.] if  $s_1, s_2, \dots , s_n, \dots $ is an
infinite sequence of elements of   $S$, then there exist integers
$i,j$  such that   $i<j$  and  $s_i \leq  s_j$.
\end{enumerate}
\end{theorem}
Let $\sigma = (s_i)_{i\geq 1}$ be an infinite sequence of
elements of a set $S$. Then $\sigma$ is called \emph{good} if it
satisfies condition ii of Theorem \ref{4TLethat} and it is called
\emph{bad} otherwise, that is, for all integers $i, j$ such that
$i < j$, $s_i \;\not\leq\; s_j$.
It is worth noting that, by condition ii above, a useful technique
to prove that $\leq$ is a wqo on $S$ is to prove that no bad
sequence exists in $S$.

For the sake of clarity, the  following well-known notions are briefly recalled.
If $u$ is a word over the alphabet $A$, then, for any $a\in A$,  $|u|_a$ denotes
the number of occurrences of $a$ in $u$. 

Given a word $v = a_1 \cdots    a_k$, with $ a_1, \dots, a_k\in A$, $v$ is
said to be a {\em subsequence (or subword)} of $u$ if there exist
words $u_1, \ldots, u_{k+1}$ such that 
$u = u_1a_1 \cdots u_k a_k u_{k+1}$.

Given two words $u, v$ over the alphabet $A$, the symbol 
$u \shuffle v$ denotes the set of words obtained by shuffle from
$u$ and $v$, that is the set of all words 
$$u_1 v_1 \cdots u_kv_k,$$
where $k\geq 1$ and $u = u_1 \cdots u_k, v = v_1 \cdots v_k$.

\section{Bad words}

In this section, we prove the ``only if'' part of Theorem
\ref{mainTheorem}.
We find convenient to split the proof
into three    sections. In the first two, we prove the claim
in the case that $w$ has one of the forms considered in 
Definition \ref{badw}.
\subsection{Words of form~\ref{form1}}

Denote by $w$
a word of the form
$$a^hb ^k, \;\;\;\;\mbox{with}\;\;h, k \geq 2,$$
and consider the sequence $(S_n)_{n\geq 1}$
of words of $A^*$ defined as: for
every $n\geq 1$,
$$S_n =
a^h(a^{2h}b^{2k})(aba^{h-1}b^{k-1})^{n}(a^{2h}b^{2k})b^k$$
\begin{proposition}\label{fsmain1}
$(S_n)_{n\geq 1}$ is a bad sequence of $\LL{w}$
with respect to $\vdash ^* _{\{w\}}$. In particular
$\vdash ^* _{\{w\}}$ is not a wqo on $\LL{w}$.
\end{proposition}
In order to prove Proposition \ref{fsmain1}, we prove some technical lemmas.
The following lemma is easily proved.
\begin{lemma}\label{fs8}
 For every $n\geq 1$, $S_n \in \LL{w}$.
    \end{lemma}

Now we recall a remarkable characterization of the words of
$\LL{w}$.  
Let  
$u$ be a word over $\{a, b\}$. Then we can  consider the following integer parameters
$$q^u _a = \lfloor  |u|_a / h \rfloor , \;\;\;\;q^u _b = \lfloor  |u|_b / k \rfloor,\;\;\;\mbox{and}$$
$$r^u _a = |u|_a\;\mbox{mod}\; h,\;\;\;\;r^u _b = |u|_b\;\mbox{mod}\; k.$$
 \begin{proposition}{\rm \cite{FS05}}
 \label{fs6}
Let $u$ be a word over the alphabet $A = \{a, b\}$. Then
$$u\in   \LL{w}$$ if and only if
the following condition holds:
$q^u _a =  q^u _b, \;\; r^u _a =  r^u _b\;$ and, for every prefix $p$ of $u$,
either $q^u _a >  q^u _b$ or $q^u _a =  q^u _b$ and $r^u _b =  0$.
 \end{proposition}

Now we recall some useful results proved in \cite{FS05}.
 \begin{definition}
 \label{defo:embed} Let $u = a_1 \cdots a_s$ and $v = b_1 \cdots
 b_t$ be two words over $A$ with $s \leq t$. An {\em embedding} of
 $u$ in $v$ is a map $f : \{1, \dots, s\}  \longrightarrow \{1, \dots, t\}$ such that $f$ is
 increasing and, for every $i = 1, \ldots, s$, $a_i = b_{f(i)}$.
 \end{definition}
  It is useful to remark that a word $u$ is a subsequence of $v$ if
 and only if there exists an embedding of $u$ in $v$.

  \begin{definition}
 \label{defo:difembed} Let $u, v \in {A}^*$ and let $f$ be an
 embedding of $u$ in $v$. Let $v = b_1 \cdots b_t$. Then $\langle v- u \rangle _f$ is the subsequence of $v$ defined as
  $$\langle v - u \rangle _f = b_{i_1} \cdots b_{i_{\ell}}$$
 where $\{i_1, i_2, \dots, i_{\ell}\}$ is the increasing sequence  of all the integers
  of  $\{1, \dots, m\}$ not belonging to $\IM (f)$.
   The word $\langle v - u \rangle _f$ is called the {\em difference
  of $v$ and $u$ with respect to $f$}.
  \end{definition}

  It is useful to remark that $\langle v - u \rangle _f$ is obtained
  from $v$ by deleting, one by one, all the letters of $u$ according
 to $f$. Moreover, an embedding $f$ of $u$ in $v$ is uniquely determined
  by two factorizations of $u$ and $v$ of the form
  $$u = a_1 a_2 \cdots a_s,\;\;\;\;\;\;\;\;
  v = v_1 a_1 v_2 a_2 \cdots v_s a_s v_{s+1}$$
  with $a_i \in A, \; v_i \in A ^*$.
  \begin{lemma}{\rm \cite{FS05}}
 \label{fs-1}
  Let $u, v \in  \LL{w}$
  such that  $u \vdash^*  _{\{w\}} v$. Then there exists an
embedding $f$ of $u$ in $v$ such that
$$\langle v - u \rangle _f \;\in\; \LL{w}.$$
  \end{lemma}

The following lemma is crucial.
 \begin{lemma}\label{fs7}
For every $i, j\geq 1$,
$$S_i \;\vdash^*  _{\{w\}}\; S_j$$
if and only if $i =j$.
    \end{lemma}
     \begin{proof}
By contradiction, suppose that the claim is false. Hence  there exist two positive integers $i<j $ such that
$S_i \vdash ^* _{\{w\}} S_j$. By Lemma \ref{fs-1}, there exists an embedding $f$ of
$S_i$ into $S_j$ such that $$\langle S_j - S_i \rangle _f\in \LL{w}.$$
We divide the proof of the lemma in the following two steps.
Let us set $$P =
a^h(a^{2h}b^{2k})(aba^{h-1}b^{k-1})^{i},$$
and remark that $P$ is a prefix of $S_i$ and $S_j$.\\
{\bf Step 1. } {\em  Let $Q = a^h(a^{2h}b^{2k})$.
The embedding $f$ is the identity on $Q$.}\\
Let us first prove that the following condition is true:
\begin{equation}\label{eqfs9}
\exists\;\;s\in\{1,\ldots, 2k\}\;\;\;\;\mbox{ with }\;\;\;\;f(3h + s) = 3h +s.
\end{equation}

By contradiction, deny. Hence we have $f( 3h + 2k) = \alpha > 3h +
2k$.  {Moreover we have $\alpha \leq |S_j| - (3k +
2h)$} since,{
otherwise, there would be no room to embed the remaining right part of
$S_i$.
Therefore, since $a^h a^{2h}b$ i s a prefix of $S_i$,}
the prefix $a^h a^{2h}$ of $Q$ must be embedded in
{a} prefix of $S_j$, {that we call $T$,}
$$T = a^h(a^{2h}b^{2k})(aba^{h-1}b^{k-1})^{L}p,$$
where
$$p\in \{a, \;aba^{h-1}\},$$
with $L\geq 0$.
Set $u = \langle T - a^h a^{2h} \rangle _f$. {Since $h, k \geq 2$, it}
 is easily checked that
$q^u _a < q^u _b$, so contradicting Proposition \ref{fs6}. Hence (\ref{eqfs9})
 is proved.

Now the previous condition  obviously implies that,  for every $s\leq 3h$,
$f(s) = s$. Consequently, if there exists a positive
integer $s$ with $1 \leq s \leq 2k$ and $f(3h +s) > 3h + s$, we would have
$$\langle S_j - S_i\rangle _f \in bA^*,$$ which contradicts
Proposition \ref{fs6}.
 Hence the embedding $f$ is the identity on $Q$.
$\diamond$

\vspace{.2cm}

\noindent
{\bf Step 2.}
{\em The embedding $f$ is the identity on
$P$.}\\
By Step 1, it suffices to prove the claim for all indexes  $s>|Q|$.
{Since $h, k \geq 2$, it} is easily checked that
$$\forall\;s=1,\ldots, h+k,\;\;\;\;f(|Q| + s) = |Q| + s.$$
Indeed, suppose that the condition above does not hold. This implies
the existence of a non empty prefix $p$ of $\langle S_j - S_i\rangle _f$
which does not satisfy Proposition \ref{fs6}.
By iterating the argument above, one completes the proof.
$\diamond$

\vspace{.2cm}

\noindent
Finally, Step 2 and the fact that
$Pa^2$ is a prefix of $S_i$
implies that
$$f(|P| + 1) > |P| + 1\;\;\;\;\mbox{or}\;\;\;\;f(|P| + 2) > |P| + 2,$$ whence
$$\langle S_j - S_i \rangle _f\in \{ab, b\}A^*,$$
which contradicts Proposition \ref{fs6}.
Hence the embedding $f$ cannot exist and thus $S_i \not{\vdash ^* _{\{w\}}} S_j$.
The proof of the lemma is thus complete.\qed

\end{proof}
Now we are able to prove the announced proposition.\\
{\bf Proof of Proposition \ref{fsmain1}:}
The claim immediately follows from Lemma \ref{fs8} and Lemma
\ref{fs7}.\qed

\subsection{Words of form~\ref{form2}}

Now denote by $w$
a word of the form
$$a^k b a^\ell b^m, \;\;\;\;
\mbox{with}\;\;k>\ell\geq1,\;\;m\geq 1.$$
and consider the sequence  $(S_n)_{n\geq 1}$ of words of $A^*$ defined as: for
every $n\geq 1$,
$$S_n = a^kba^\ell a^{k}b a^{\ell }b^{m} (a^k b^{m+1}a^\ell)^n
a^{k}b a^{\ell }b^{m} b^m.$$
We prove the following result.
\begin{proposition}\label{fsmain}
$(S_n)_{n\geq 1}$ is a bad sequence of $\LL{w}$
with respect to $\vdash ^* _{\{w\}}$. In particular
$\vdash ^* _{\{w\}}$ is not a wqo on $\LL{w}$.
\end{proposition}
The following lemma is  easily proved.
\begin{lemma}\label{fs3}
 For every $n\geq 1$, $S_n \in \LL{w}$.
    \end{lemma}

Let us define the map $\nu : A^+ \;\longrightarrow\; \Q \;\cup\;\{\infty\},\;$
as: for every $u\in A^*$,
$$\nu (u) = \frac{|u|_a}{|u|_b}.$$
The following two lemmas are easily proved by
induction on the length of the derivation used to
obtain $u$.
\begin{lemma}\label{fs1}
Let $u\in \LL{w}$. For every  non empty prefix $p$ of
$u$, we have
$$\nu (p) \geq \frac{k+\ell}{m+1}.$$
    \end{lemma}
\begin{lemma}\label{fs2}
Let $u$ be a word of $\LL{w}$.
If $a^\alpha b$ is a prefix of $u$, then $\alpha \geq k$.
 If $a^\alpha b^2$ is a prefix of $u$, then $\alpha \geq 2k$.
    \end{lemma}
The following lemma is crucial.

\begin{lemma}\label{fs4}
For every $i, j\geq 1$,
$$S_i \;\vdash^*  _{\{w\}}\; S_j$$
if and only if $i =j$.
    \end{lemma}
     \begin{proof}
By contradiction, suppose that the claim is false. Hence  
there exist two positive integers $i<j $ such that
$S_i \vdash^*  _{\{w\}}S_j$. By Lemma \ref{fs-1}, there exists an embedding $f$ of
$S_i$ into $S_j$ such that $$\langle S_j - S_i \rangle _f\in \LL{w}.$$
We divide the proof of the lemma in the following two steps.
Let us set $$P = a^kba^\ell a^{k}b a^{\ell }b^{m},$$
and remark that $P$ is a prefix of $S_i$ and $S_j$.

\vspace{.2cm}

\noindent
{\bf Step 1. } {\em  The embedding $f$ is the identity on $P$.}\\
Set
$Q = a^k b a^\ell a^{k} b.$
We first show that:
\begin{equation}
\label{eqfs5}
\exists\;\; s\in \{1, \ldots, \ell \}, \;\;\;\;\;\;\;\;\mbox{where}\;\; f(|Q| + s) = |Q| + s.
\end{equation}
By contradiction, suppose that (\ref{eqfs5}) does not hold.
Consequently $f(|P|) > |P|$.
Since $a^{k}b a^{\ell }b^{m}b^m$ is a suffix of $S_i$,
$f(|P|) < |P(a^kb^{m+1}a^\ell)^j|.$
Since $P$ ends with $b$ and $Pa$ is a prefix of $S_i$,
the prefix $P$ of $S_i$ must be embedded (according to $f$)
in a prefix of $S_j$, we call $T$,
$$T = Pa^k (b^{m+1} a^{\ell + k})^\beta b^{m+1},$$
where $\beta$ is such that $0 \leq \beta < j$.
Therefore, the word $\langle T - P\rangle _f$
is a prefix of $\langle S_j - S_i\rangle _f$.
On the other hand, an easy computation shows that
$$\nu (\langle T - P\rangle _f) =
\frac{|\langle T - P\rangle _f|_a}{|\langle T - P\rangle _f|_b} =
\frac{\beta (\ell + k)  + k }{(1 + \beta ) (m+1)},$$
and thus
$$\nu (\langle T - P\rangle _f) < \frac{k+\ell}{m+1},$$ so
contradicting Lemma \ref{fs1}. Thus condition (\ref{eqfs5}) is proved:
it means that $f$ is the identity on $Q$. Finally this
condition implies that $f$ is the identity on $P$. 
Indeed, otherwise,
$\langle S_j - S_i \rangle _f \;\in a^\alpha bA^*,$
with $0 \leq \alpha < l$ which contradicts
Lemma \ref{fs2} since $l < k$.
$\diamond$

\vspace{.2cm}

\noindent
{\bf Step 2.}
{\em The embedding $f$ is the identity on
$P(a^k b^{m+1}a^\ell)^i$.}\\
By Step 1, it suffices to prove the claim for all indexes  $s>|P|$.
It is easily checked that, for every $s = 1, \ldots, m+1+\ell+2k$,
$$f(|P| + s) = |P| + s.$$ Indeed, otherwise,
we would have $\langle S_j - S_i \rangle _f \;\in a^\alpha b^2A^*,$
with $\alpha < 2k$ or  $\langle S_j - S_i \rangle _f \;\in a^\alpha bA^*,$
with $\alpha < k$, so contradicting  Lemma \ref{fs2}.
 By iterating the argument above, one completes the proof.
$\diamond$

\vspace{.2cm}

\noindent
We have already proved that $S_i = P'R$, $S_j =
P'(a^kb^{m+1}a^\ell)^{j-i}R$ where $P' = P(a^kb^{m+1}a^\ell)^i$ and $R
= a^{k}b a^{\ell }b^{m}b^m$, and that $f$ is the identity on $P'$.
It follows that $\langle S_j - S_i \rangle_f$ begins with a prefix which is 
$a^kb^2$ (if $f(|P'|+1) > f(|P'| + k + m + 1)$) or 
$a^\alpha b$ where $\alpha < k$ so contradicting Lemma \ref{fs2}.
Hence the embedding $f$ cannot exist and thus $S_i \not{\vdash^*  _{\{w\}}} S_j$.
The proof of the lemma is thus complete.\qed
\end{proof}

Now we are able to prove the announced proposition.\\
{\bf Proof of Proposition \ref{fsmain}:}
The claim immediately follows from Lemma \ref{fs3} and Lemma
\ref{fs4}.\qed

\subsection{Proof of the ``only if'' part of Theorem~\ref{mainTheorem}}

As pointed out in the previous paragraph, Propositions \ref{fsmain1} and \ref{fsmain} permit to prove that
if $w$ is of the forms (1) or (2) of Definition \ref{badw}, then
$\vdash^* _{\{w\}}$ is not a wqo on $\LL{w}$. This does not suffice
to prove the ``only if'' part of Theorem \ref{mainTheorem}. In order to complete
the proof, the following lemma (and its symmetric version, say
Lemma \ref{fs0bis}) provides a key result: indeed it shows that the 
property ``$\vdash^*  _{\{w\}}$ is not a wqo on $\LL{w}$''
is preserved by the factor order.

    \begin{lemma}\label{fs0}
    Let $b$ be a letter of an alphabet $A$ and let $u$ be a word over $A$
not ending with $b$. Assume $\vdash^*_{\{u\}}$ is not a wqo on
$\LL{u}$. Then,  for every $k\geq 1$,
$\vdash^*_{\{ub^k\}}$ is not a wqo on $\LL{ub^k}$.
    \end{lemma}
    \begin{proof}
Let $(w_n)_{n\geq 0}$
be a bad sequence of  $\LL{u}$ with respect to
$\vdash^*_{\{u\}}$ and,
  for every $n\geq 1$, let us denote $\ell _n$
the positive  integer such that
\begin{equation}\label{eqfs0}
\epsilon \;\vdash^ {\ell _n} _{\{u\}} \; w_n.
\end{equation}
Since $(w_n)_{n\geq 0}$ is a bad sequence,
by using a standard argument, we may choose the sequence $(w_n)_{n\geq 0}$
so  that $(\ell _n)_{n\geq 0}$  is a strictly increasing
sequence of positive integers.
Let $k$ be a positive integer and define the sequence of words
$(w_n(b^k)^{\ell _n})_{n\geq 0}$.
It is easily checked that, for every $n\geq 1$,
$$\epsilon \;\vdash^ {\ell _n} _{\{ub^k\}} \; w_n(b^k)^{\ell _n},$$
so that all the words of the sequence defined above belong to
the language $\LL{u}$.
Now we prove that this sequence is bad
with respect to
$\vdash^*_{\{ub^k\}}$.
By contradiction, suppose the claim false. Thus there exist positive integers $n, m$  such that
\begin{equation}\label{eqfs1}
w_n(b^k)^{\ell _n}\;\vdash^ {*} _{\{ub^k\}} \; w_{n+m}(b^k)^{\ell _{n+ m}}.
\end{equation}
Since, for every $n\geq 1$,
$$|w_n(b^k)^{\ell _n}| =  \ell _n k + |w_n|= \ell_n(k + |u|),$$
we have that the length $L$ of  the derivation
(\ref{eqfs1}) is
\begin{equation}\label{eqfs2}
L = \ell _{n+ m} - \ell _n.
\end{equation}

Now it is useful to  do the following remarks. First observe that, since $u$ does not end with the letter $b$,
for every $n\geq 1$, $(b^k)^ {\ell_n}$ is the longest power of $b$ which is a  suffix of
$w_n (b^k)^{\ell_n}$.
Second: at each step
$$ v \;\vdash _{\{ub^k\}} \; v',$$
of the derivation  (\ref{eqfs1}), the exponent of the
longest power of $b$ which is a suffix of
the word  $v'$ increases of $k$ at most
(with respect to $v$).
Moreover this upper bound can be  obtained by performing the insertion of
$ub^k$ in the word $v$
only if its suffix $b^k$  is
inserted after the last letter of $v$ {which is different from $b$}.
By the previous remark and by
(\ref{eqfs2}), all the insertions of the derivation (\ref{eqfs1}) must be done
in this way. This implies that
the derivation
(\ref{eqfs1}) defines in an obvious way a new one with respect to
the relation  $\vdash^ {*} _{\{u \}}$ such that
$$
w_n \;\vdash^ {*} _{\{u \}} \; w_{n+\ell}.
$$
The latter condition contradicts the fact that
the sequence of words $(w_n)_{n\geq 0}$  is bad.
 \qed
    \end{proof}
By using a symmetric argument, we can prove the following.
  \begin{lemma}\label{fs0bis}
    Let $b$ be a letter of an alphabet $A$ and let $u$ be a word over $A$
not beginning with $b$. Assume $\vdash^*_{\{u\}}$ is not a wqo on
$\LL{u}$. Then,  for every $k\geq 1$,
$\vdash^*_{\{b^ku\}}$ is not a wqo on $\LL{b^ku}$.
    \end{lemma}
We are now able to prove the sufficiency of Theorem \ref{mainTheorem}.
\begin{theorem}
\label{fscondsuf}
If $w$ is a bad word
then $\vdash ^* _ {\{w\}}$ is not a
wqo on the language $\LL{w}$.
\end{theorem}
\begin{proof}
If $w$ has a factor of the form
$a^kb^h$ {with} $h, k\geq 2,$ or
$a^k b a^\ell b^m,$ {with} $k > \ell\geq 1, \; m\geq 1,$
then the claim is
a straightforward consequence of Lemma \ref{fs0},
Lemma \ref{fs0bis}, Proposition \ref{fsmain1}, and Proposition \ref{fsmain}.

In the general case, that is whenever $\tilde w$ or $E(w)$ or $E(\tilde w)$
has a factor of the previous two  forms, the proof is similar since the
property of wqo is preserved under taking exchange morphism and mirror
image of the word $w$.
\qed
\end{proof}
\section{Good words}
In this section we present the proof of the
``if''  part of Theorem~\ref{mainTheorem}. We find convenient to
split it into the following seven sections. In the first a
characterization of good words
and that  of the languages
of words derivable from  a good word
are given.
\subsection{Form of good words}

\begin{lemma}
\label{goodWords}
A  word $w$ is good if and only if $w = \epsilon$ or there
exist some integers $n, e, i, f$ such that $w = a^i(ba^n)^eba^f$ or $w
= b^i(ab^n)^eab^f$, $e \geq 0$, $0 \leq i, f \leq n$, and if $e = 0$
then $n = \max(i, f)$.
\end{lemma}

\begin{proofWithQed}
Clearly if $w$ is a bad word, then $w$ cannot be decomposed as in the
lemma.

Assume now that $w$ is a good word. This means that $w$ has no factor
of the form $aabb$, $bbaa$, $a^{n+1}ba^nb$, $ba^nba^{n+1}$,
$b^{n+1}ab^na$, $ab^nab^{n+1}$ with $n \geq 1$ an integer.

If $|w|_a = 0$, then $w = \epsilon$ or $w = a^i(ba^n)^eba^f$ with
$i = n = f = 0$.  If $|w|_a = 1$, $w = a^pba^q$  with
$\max (p,q) =1$, that is $w = a^i(ba^n)^eba^f$ with $i = p$, $f = q$, $n =
\max(p,q)$, $e = 0$.  Similarly if $|w|_b \leq 1$, $w$ is a good word.

Assume from now on that $|w|_a \geq 2$ and $|w|_b \geq 2$. If both
$aa$ and $bb$ are not factors of $w$, then $w$ is a factor of
$(ab)^\omega$ and so $w = a^i(ba^n)^eba^f$ with $n = 1$.

Let us prove that $aa$ and $bb$ cannot be simultaneously factors
of $w$. Assume the contrary. We have $w = w_1aaw_2bbw_3$ (or $w =
w_1bbw_2aaw_3$ which leads to the same conclusion) for some words
$w_1, w_2, w_3$.  Without loss of generality we can assume that $aa$
is not a factor of $aw_2$ and $bb$ is not a factor of $w_2b$. This
implies that $w_2 = (ba)^m$ for an integer $m \geq 0$. This is not
possible since $aabab$ and $aabb$ are not factors of $w$.

Assume from now on that $bb$ is not a factor of $w$ (the case where
$aa$ is not a factor is similar). This implies that $w =
a^{i_0}ba^{i_1}ba^{i_2}b\ldots ba^{i_p}ba^{i_{p+1}}$ for some integers
$i_0$, $i_1$, \ldots, $i_{p+1}$ such that $i_j \neq 0$ for each $j \in
\{1, \ldots, p\}$. Let $j$ be an integer such that $1 \leq j < j+1
\leq p$. Since $a^{i_{j+1}+1}ba^{i_{j+1}}b$ and $ba^{i_j}ba^{i_j+1}$
are not factors of $w$, we have $i_j = i_{j+1}$. Thus set
$n = i_1 = \cdots = i_p$ and write $w =
a^{i_0}(ba^n)^pba^{i_{p+1}}$. Since $a^{n+1}ba^nb$ and $ba^nba^{n+1}$
are not factors of $w$, we have $i_0 \leq n$, $i_{p+1} \leq n$. This
ends the proof.
\end{proofWithQed}

\medskip

For $X$ a set of words and $n$ an integer, let $X^{\leq n} =
\bigcup_{i = 0}^n X^i$. Then  Lemma~\ref{goodWords} can be reformulated:
the set of good words $w$ is the set
$$\{\epsilon\} \cup \bigcup_{n \geq 0} a^{\leq n}(ba^n)^*ba^{\leq n}
\cup \bigcup_{n \geq 0} b^{\leq n}(ab^n)^*ab^{\leq n}.$$

\subsection{A fundamental characterization}

In this section we prove the next proposition that characterizes words
in $\LL{w}$ when $w$ is a good word.  The construction which is made
in order to prove it also allows us to prove $\vdash ^* _{\{w\}}$'s
properties (see Lemma~\ref{propertiesNumbering}) on some prefixes of
elements of $\LL{w}$.

\begin{proposition}

\label{carac} Let $w$ be a word over $\{a, b\}$ and let  $n_w, e_w, i_w,
f_w$ be integers such that $|w|_a \geq 1$, $|w|_b \geq 1$, $w =
a^{i_w}(ba^{n_w})^{e_w}ba^{f_w}$, where $0 \leq i_w, f_w \leq n_w$, $e_w
\geq 0$ and if $e_w = 0$ then $n_w = \max(i_w, f_w)$.

A word $u$   belongs to $\in \LL{w}$ if and only if the following 
conditions are satisfied:
\begin{enumerate}
\item ${|u|_a \over |w|_a} = {|u|_b \over |w|_b}$;
\item for all words $p$, $s$, if $u = ps$ then
   \begin{description}
   \item{2.1)} $|p|_a \geq i_w |p|_b + \max(0, |p|_b -
{|u|_b \over |w|_b}) (n_w - i_w)$;
   \item{2.2)} $|s|_a \geq f_w |s|_b + \max(0, |s|_b - {|u|_b \over
   |w|_b}) (n_w - f_w)$.
   \end{description}
\end{enumerate}
\end{proposition}

In order to prove Conditions 1, 2.1 and 2.2, we now
introduce a numbering of the letters which has very good properties
(see in particular Lemma~\ref{propertiesNumbering}) when the word
verifies the three conditions above.

Let $w$, $n_w$, $e_w$, $i_w$ and $f_w$ be as in Proposition~\ref{carac}.
Let $u$ be a word verifying Condition~1 of Proposition~\ref{carac}
and let $x = {|u|_a \over |w|_a} = {|u|_b \over |w|_b}$.
We observe that if $u \in \LL{w}$ then $u$ is the shuffle of $x$ occurrences of $w$.

For any $\alpha \in \{a, b\}$, let $\pi_\alpha$ be the function defined on
$\{1, \ldots, |u|_\alpha\}$ as follows: $\pi_\alpha(i)$ is the index of the
$i^{\rm th}$ occurrence of the letter $\alpha$ in $u$.

\noindent
\textit{Example.} Let $w = abaaabaa$ and let $u =
abaaababaabaaabaabaaaaabaaaabaaa$. We have $x = 4$, 
$\pi_b(1) = 2$, $\pi_b(2) = 6$, 
$\pi_b(3) = 8$, $\pi_b(4) = 11$, 
$\pi_b(5) = 15$, $\pi_b(6) = 18$, 
$\pi_b(7) = 24$, $\pi_b(8) = 29$.

\medskip

In order to find $x$ occurrences of $w$ in $u$, for every $1 \leq i \leq x$, 
we  define the following set of integers:

\begin{eqnarray*}
P(i) &=& \{ \pi_a( (i-1)i_w + j ) \mid 1 \leq j \leq i_w \} \\ 
&&  \cup ~
\{ \pi_a( x i_w + k x n_w + (i-1) n_w + j ) \mid 1 \leq j \leq n_w, 0 \leq k <
e_w\} \\ 
&& \cup ~\{ \pi_a( x i_w + e_w x n_w + (i-1) f_w + j ) \mid 1 \leq
j \leq f_w\} \\
&&  \cup  ~ \{ \pi_b( i + k x ) \mid 0 \leq k \leq e_w \}.\\ 
\end{eqnarray*}

Note that the idea for introducing the sets $P(i)$ is to try to mark
(when $u \in \LL{w}$) some possible occurrences of $w$ as subsequences
of $u$ (see also words $u(i)$ below).

\noindent
\textit{Example (continued).} We have~:\\
\begin{eqnarray*}
P(1) &=& \{ 1, 2, 7, 9, 10, 15, 23, 25 \},\\
P(2) &=& \{ 3, 6, 12, 13, 14, 18, 26, 27\},\\
P(3) &=& \{ 4, 8, 16, 17, 19, 24, 28, 30\},\\ 
P(4) &=& \{ 5, 11, 20, 21, 22, 29, 31, 32\}.\\
\end{eqnarray*}

The following properties easily follow from the definition of the sets
$P(i)$ above:
\begin{enumerate}
\item The family $\{P(i)\}_{¥1 \leq i \leq x}\;$ is  a partition of the set $\{1, \ldots, |u|\}$.
\item For each $i$ with $1 \leq i \leq x$, the set $P(i)$ has exactly
$|w|$ elements.
\end{enumerate}

Let $i$ be an integer with $1 \leq i \leq x$. Assume that $P(i) = \{i_1, \ldots,
i_{|w|}\}$ with $i_1 < i_2 < \ldots < i_{|w|}$. We denote by $u(i)$ the
word $u_{i_1}u_{i_2} \ldots u_{i_{|w|}}$. (In the example, $u(1) = 
u_1u_2u_7u_9u_{10}u_{15}u_{23}u_{25} = abaaabaa = w$).

Let us observe that, from an intuitive point of view,
it could be useful  to consider the word over the alphabet $\{1, \ldots, x\}$ 
defined as follows: for any $i \in \{1, \ldots, |u|\}$,
the $i^{\rm th}$ letter of the word is the integer $j$ such that $i
\in P(j)$.
\medskip 

\noindent
\textit{Example (continued).} 
In the first row, we write the word $u$, while in the second, we write
the word defined above:

\medskip 
\centerline{\begin{tabular}{l}
\tt abaaababaabaaabaabaaaaabaaaabaaa\\
\tt 11234213114222133234441312234344.\\
\end{tabular}
}

\medskip

Some useful properties of the previous numbering are proved in the
next lemma.

\begin{lemma}
\label{propertiesNumbering}
Let $w$ (resp. $u$) be a word verifying the hypotheses
(resp. Conditions 1 and 2) of Proposition~\ref{carac}. Let $x = {|u|_a
\over |w|_a}$. Then the following conditions hold:

\begin{enumerate}
\item For each $1 \leq i \leq x$, $u(i) = w$. Consequently, $u \in
\LL{w}$.

\item If $p$ is a prefix of $u$ such that $|p|_a = x (i_w + kn_w)$
with $0 \leq k \leq e_w$, then $p \in \LL{p_{w,k}, p_{w,k}b}$ where
$p_{w, k} = a^{i_w}(ba^{n_w})^k$.
\end{enumerate}
\end{lemma}

\begin{proofWithQed}
Let $i$, $1 \leq i \leq x$. The fact that $u(i) = w$ follows immediately 
the definition of $u(i)$ (and $P(i)$) and the three following properties~:

\begin{description}
\item {\it Property 1.} If $p$ is a word such that $pb$ is a prefix of
$u$ and $|pb|_b = i$ then $|p|_a \geq i_w
|pb|_b = i_w \times i$. This shows that $\pi_a(i_w (i-1) + j) <
\pi_b(i)$ for each $1 \leq j \leq i_w$.

\textit{Proof of Property~1.} 
By Condition~2.1 of Proposition~\ref{carac}, $|p|_a = |pb|_a \geq i_w
|pb|_b$.

\item{\it Property 2.}  If $p$ and $s$ are the words such that $u =
pbs$ and $|pb|_b = e_w x + i$ (that is $|s|_b = x - i$) then $|p|_a
\leq x (i_w + e_w n_w) + (i-1) f_w$. This shows that $\pi_b(e_w x + i) <
\pi_a( x i_w + e_w x n_w + (i-1) f_w + j)$ for each $1 \leq j \leq i_w$.

\textit{Proof of Property~2.}  
By Condition~2.2 of Proposition~\ref{carac},
$|s|_a = |bs|_a \geq f_w |bs|_b$.  Since $|u|_a = |s|_a + |p|_a$ and
$|u|_a = x(i_w + n_w e_w + f_w)$, $|p|_a \leq x(i_w + n_w e_w) + f_w
(x - |bs|_b) =  x (i_w + n_w e_w) + f_w (i-1)$.

\item{\it Property 3.} 
If $p, v, s$ are the words such that $u = pbvbs$ with $|pb|_b = i + kx$
with $0 \leq k < e_w$, and $|pbvb|_b = i + (k+1)x$, then $|pb|_a \leq x
i_w + (kx + i-1)n_w$ and $x i_w + (kx + i)n_w \leq |pbvb|_a$. This means
that $|pb|_b = \pi_b(i + kx) < \pi_a(x i_w + (kx + i-1)n_w +j)
< \pi_b(i+(k+1)x)$ for each $1 \leq j \leq n_w$.

\textit{Proof of Property~3.}  First we observe that $|pbvb|_b > x$
and so $\max(0, |pbvb|_b - x) = |pbvb|_b -x$. Hence by Condition~2.1
of Proposition~\ref{carac}, $|pbvb|_a \geq i_w |pbvb|_b + (|pbvb|_b
- x) (n_w - i_w) = i_w x + n_w (|pbvb|_b-x) = i_w x + n_w(i + k x)$.

Now we observe that $|bvbs|_b \geq x$ (Indeed $|bvbs|_b = |u|_b - |p|_b = x(e_w
+ 1) - (i + k x - 1) = x + x(e_w-k-1) + (x-i+1) > x$) and so $\max(0,
|bvbs|_b -x) = |bvbs|_b -x$. Hence by Condition~2.2 of
Proposition~\ref{carac}, $|bvbs|_a \geq f_w |bvbs|_b + (|bvbs|_b -x)
(n_w- f_w) = f_w x + n_w (|bvbs|_b -x)$.  Since $|u|_a = |p|_a +
|bvbs|_a$ and $|u|_a = x(i_w + n_w e_w + f_w)$, we have $|p|_a \leq x
i_w + n_w (x e_w + x - |bvbs|_b)$.  But $(e_w+1)x = |u|_b = |p|_b +
|bvbs|_b = i + kx - 1 + |bvbs|_b$, that is $x e_w + x - |bvbs|_b = i +
kx -1$.  Thus $|pb|_a = |p|_a \leq x i_w + (kx + i -1)n_w$.
\end{description}

Let us now prove the second part of Lemma~\ref{propertiesNumbering}.

First we observe that $xk \leq |p|_b \leq x(k+1)$.  Indeed if $|p|_b <
xk$, then considering the word $s$ such that $u = ps$, $|s|_b >
x(e_w+1-k) \geq x$, and by Condition~2.2 of Proposition~\ref{carac},
$|s|_a \geq f_w |s|_b + (|s|_b - x) (n_w-f_w) > f_w x (e_w+1-k) + x
(e_w-k) (n_w-f_w) = x (f_w + (e_w-k) n_w)$, and so $|p|_a = |u|_a -
|s|_a = x(i_w + e_w n_w + f_w) - |s|_a < x(i_w + k n_w)$ which
contradicts the hypotheses. Moreover if $|p|_b > x(k+1) \geq x$, by
Condition~2.1, $|p|_a \geq i_w |p|_b + (|p|_b - x) (n_w - i_w) > i_w
x(k+1) + kxn_w - kx i_w = x(i_w+k n_w)$ which also contradicts the hypotheses.

Let $p(i)$, $1 \leq i \leq x$, be the prefix of $u(i)$ constituted of
the letters with index in $P(i) \cap \{1, \ldots, |p|\}$.  From $xk
\leq |p|_b \leq x(k+1)$, we deduce that the set $\{ \pi_b(i+ \ell x)
\mid 0 \leq l < k\}$ is included in the set $P(i) \cap \{1, \ldots,
|p|\} \cap \{\pi_b(j) \mid 1 \leq j \leq |u|_b\}$ which itself is
included in the set $\{ \pi_b(i+ \ell x) \mid 0 \leq l \leq
k\}$. Hence $k \leq |p(i)|_b \leq k+1$.  Moreover since $|p|_a = x(i_w
+ k n_w)$, the set $P(i) \cap \{1, \ldots, |p|\} \cap \{\pi_a(j) \mid
1 \leq j \leq |u|_a\}$ equals the union of the sets
$\{\pi_a((i-1)i_w+j) \mid 1 \leq j \leq i_w\}$ and $\{\pi_a(xi_w+
\ell x n_w + (i-1) n_w + j) \mid 1 \leq j \leq n_w, 0 \leq \ell < k\}$,
so that
$|p(i)|_a = i_w + k n_w$.  Since $u(i) = w$, we deduce that $p(i) \in
\{p_{k,w}, p_{k,w}b\}$ and so $p \in \LL{p_{k,w}, p_{k,w}b}$.
\end{proofWithQed}

\medskip

\begin{proof2}{Proposition~\ref{carac}}
The sufficiency of Conditions~1 and 2  is ensured by
Lemma~\ref{propertiesNumbering} (1).
\medskip

It is immediate that Condition~(1) is necessary. 
We prove that it is also the case for Condition~2.1, the proof for
Condition~2.2 being similar.  
Let $u \in \LL{w}$ and let $x$ be the integer such that
$\epsilon \qo{w}^x u$. 
If $x = 0$ then $u=\epsilon$ and the claim is trivially verified.
Thus suppose $x>0$.

We have $|u|_a = x |w|_a$ and $|u|_b = x |w|_b$, 
so that  $x = |u|_b /|w|_b = |u|_a /|w|_a$.

Since $u$ is the shuffle of $x$ occurrences of $w$, any prefix $p$ of
$u$ is the shuffle of $x$ prefixes of $w$: there exist prefixes $p_1$,
\ldots, $p_x$ such that $$p\in p_{1}\;\shuffle\;\cdots\;\shuffle\;
p_{x},$$
Thus 
$$|p|_{a} = \sum _{i=1, \ldots, x}\;|p_{i}|_{a}.$$ Since $p_i$ is a
prefix of $w$, if $|p_i|_b \neq 0$, $|p_i|_a \geq i_w+(p_i|_b-1)n_w$.
Assume without loss of generality that $p_1, \ldots, p_{x'}$ contain
at least one $b$ and that $p_{x'+1}, \ldots, p_x$ contain no $b$.  We get

$$|p|_{a}\geq x'i_{w} + n_{w} \sum_{i=1, \ldots, x'}\;|p_i|_{b} - x'n_w.$$
But $|p|_{b} = \sum_{i=1, \ldots, x'}\;|p_i|_{b}$. So


$$|p|_{a}\geq x'i_{w} + n_{w}(|p|_{b}-x') = i_{w}|p|_{b} +
(|p|_{b}-x')(n_{w} -  i_{w}).$$
Since $x' \leq x = |u|_b /|w|_b$, the latter inequality gives
$$|p|_a \geq i_w |p|_b + \max(0, |p|_b -
{|u|_b \over |w|_b}) (n_w - i_w).$$
 The proof is thus complete.
\end{proof2}

\subsection{Some useful wqo's}

In this section, we present some useful wqo's. First
we recall the following result.

\begin{proposition}{\rm \cite{FS05}}
\label{propFS05}
For any integer $n \geq 0$,
if $w \in \{a^nb, ab^n, ba^n, b^na \}$,
$\qoEtoile{w}$ is a wqo on
$\LL{w} = {\rm L}_{{w}}^{\epsilon}$.
\end{proposition}

This result allows us to state:

\begin{lemma}
\label{lemmaUsefulWQO1}
Let $n \geq 0$ be an integer.  Let $I$ be one of the following sets:
$\{a^nb, a\}$, $\{a^nb, b\}$, 
$\{b^na, a\}$, $\{b^na, b\}$,
$\{ba^n, a\}$, $\{ba^n, b\}$, 
$\{ab^n, a\}$, $\{ab^n, b\}$:
$$\LLSet{I} = {\rm L}_{{I}}^{\epsilon}.$$
\end{lemma}

\begin{proofWithQed}Assume $I = \{a^nb, a\}$.
It is immediate that ${\rm L}_{{I}}^{\epsilon} \subseteq \LLSet{I}$.  Let
$w$ be a word in $\LLSet{I}$. There exists a word $w_1$ such that
$\epsilon \qoEtoile{a^nb} w_1 \qoEtoile{a} w$. By
Proposition~\ref{propFS05}, $w_1 \in {\rm L}_{{a^nb}}^{\epsilon}$, and
so $w \in {\rm L}_{{I}}^{\epsilon}$.

 The proof
for the other values of $I$ is similar.
\end{proofWithQed}

\begin{lemma}
\label{lemmaUsefulWQO2} Let $n \geq 1$ be an integer.
The three following assertions are equivalent for a word $w$:
\begin{enumerate}
\item $w \in \LL{a^nb, a^n}$;
\item $|w|_a = 0 \mod n$,
and, for any prefix $p$ of $w$, $|p|_a \geq n |p|_b$;
\item $w \in {\rm L}_{\{a^nb, a^n\}}^\epsilon$.
\end{enumerate}
In particular, $\LL{a^nb, a^n} = {\rm L}_{\{a^nb, a^n\}}^\epsilon$.
\end{lemma}

\begin{proofWithQed}
$3 \Rightarrow 1$ is immediate.

For any word $w$ in $\LL{a^nb, a^n}$, obviously $|w|_a = 0 \mod
n$. Moreover $w$ is a prefix of a word in 
$\LL{a^nb}$. Thus $1 \Rightarrow 2$ is a
direct consequence of Proposition~\ref{carac}. Indeed taking $w = a^nb$, $n_w
= n = i_w$, and $e_w = f_w = 0$, Condition~2.1 of Proposition~\ref{carac}
says that for any prefix of a word in $\LL{a^nb}$, $|p|_a \geq i_w|p|_b
= n|p|_b$.

We now prove $2 \Rightarrow 3$ by induction on $|w|_b$.  Since $|w|_a
= 0 \mod n$, the result is immediate if $|w|_b = 0$.  Assume $|w|_b \geq
1$.  Assertion~2 on $w$ implies the existence of an integer $k \geq 0$ and a
word $w'$ such that $w = a^k a^n b w'$.  Let $p$ be a prefix of
$a^kw'$.  If $|p| \leq k$, then $n |p|_b = 0 \leq |p|_a$.  If $|p| >
k$, $p = a^kp'$ for a prefix $p'$ of $w'$. Assertion~2 on $w$ implies
that $|a^ka^nbp'|_a \geq n |a^ka^nbp'|_b$ that is $|a^kp'|_a \geq
n|a^kp'|_b$. Thus $a^kw'$ verifies Assertion~2 and so by inductive
hypothesis, $a^nw' \in {\rm L}_{\{a^nb, a^n\}}^\epsilon$. It follows that
$w \in {\rm L}_{\{a^nb, a^n\}}^\epsilon$.
\end{proofWithQed}

\medskip

Similarly to Lemma~\ref{lemmaUsefulWQO2}, one can state that $\LL{ba^n,
a^n} = {\rm L}_{\{ba^n, a^n\}}^\epsilon$ (this needs to exchange
prefixes by suffixes), and, exchanging the roles of $a$ and $b$,
$\LL{b^na, b^n} = {\rm L}_{\{b^na, b^n\}}^\epsilon$ and
$\LL{ab^n, b^n} = {\rm L}_{\{ab^n, b^n\}}^\epsilon$.

\medskip

Let us recall that:
\begin{theorem} {\rm \cite{DV2003,DV2004}}
\label{resDV1}
For any finite set $I$, $\vdash_{I}^*$ is a wqo on
${\rm L}_{{I}}^{\epsilon}$.
\end{theorem}

Hence from this theorem and the previous lemma, we deduce:

\begin{proposition}
\label{usefulWQO}
Let $n \geq 0$ be an integer.  Let $I$ be one of the following sets:
$\{a^nb, a\}$, $\{a^nb, b\}$, 
$\{b^na, a\}$, $\{b^na, b\}$,
$\{ba^n, a\}$, $\{ba^n, b\}$, 
$\{ab^n, a\}$, $\{ab^n, b\}$,
$\{a^nb, a^n\}$,
$\{ba^n, a^n\}$,
$\{b^na, b^n\}$,
$\{ab^n, b^n\}$. The derivation relation $\qoEtoileSet{I}$
is a wqo on $\LLSet{I}$.
\end{proposition}

\subsection{A decomposition tool}

\begin{lemma}
\label{basicLemmaGeneralized}
Let $m \geq 1$ be an integer.
Any word $w$ over $\{a, b\}$ can be factorized as $w = w_1w_2w_3$ 
with $w_1 \in \LL{ba^m, a}$, $w_2 \in \LL{ba^m, b}$ and $|w_3|_a <m$.

Moreover, if $w$ is the shuffle of $x$ occurrences of $ba^m$ and of a
word $w'$, then $x \leq |w_1|_{b} + |w_2|_a/m$.
\end{lemma}

\begin{proofWithQed}
We prove the first part of this result by induction on $|w|$. The 
claim is
trivial if $w = \epsilon$. Assume $|w| \geq 1$, so that  $w =
w'\alpha$ with $\alpha \in \{a, b\}$. By inductive hypothesis, $w' =
w_1'w_2'w_3'$ with $w'_1 \in \LL{ba^m, a}$, $w'_2 \in \LL{ba^m, b}$
and $|w'_3|_a <m$.

If $\alpha = b$ or if $\alpha = a$ and $|w'_3\alpha|_a <m$, the
result is true for $w$ by setting  $w_1 = w_1'$, $w_2 = w_2'$ and $w_3 =
w_3'\alpha$.  Assume now that $\alpha = a$ and $|w'_3\alpha|_a =
m$. Two cases have to be considered. If $w_2' \not\in \LL{ba^m}$, then
$w_2'w_3'a \in \LL{ba^m, b}$ and thus we can set $w_{1}=w' _{1}$,
$w_{2 }= w_2'w_3'a$ and $w' _{3}=\epsilon$.

Consider now that $w_2' \in \LL{ba^m}$. By
replacing $w_1'$ (resp. $w_2'$) by $w_1'w_2'$ (resp. $\epsilon$), we
can assume $w_2' = \epsilon$. If $w_3'$ starts with $b$, then $w_3'a
\in \LL{ba^m, b}$ and the result is true for $w$ with $w_1 = w_1'$, $w_2 = w_2'w_3'$ 
and $w_3 = \epsilon$. If $w_3'$ starts with $a$, $w_3' = a x$ for a
word $x$. The result is true for $w$ with $w_1 = w_1'a$, $w_2 =
w_2' = \epsilon$ and $w_3 = x$.

The argument used in the induction above can be
used  for the proof of the
second part of the statement of Lemma~\ref{basicLemmaGeneralized}.
\end{proofWithQed}

\subsection{A first inductive result}

The aim of this section is to prove the next result which proof is
based on the characterization provided by
Proposition~\ref{inductiveDecomposition1}.

\begin{proposition}
\label{inducStep1}
Let $n, m$ be two integers such that $n, m \geq 1$ and let $w$ be a
 word in $a^{\leq n}(ba^n)^*b \cup \{\epsilon\}$ such
 that $wa^nba^m$ is a good word.  If $\qoEtoile{wa^n, wa^nb}$ is a wqo
 on $\LL{wa^n, wa^nb}$ then $\qoEtoile{wa^nb, wa^nba^m}$ is a wqo on
 $\LL{wa^nb, wa^nba^m}$.
\end{proposition}

Observe that the hypothesis ``$wa^nba^m$ is a good word'' means only 
$1 \leq m \leq n$ when $w \neq \epsilon$.

\begin{proposition}
\label{inductiveDecomposition1}
Let $n, m$ be two integers such that $n, m \geq 1$ and let $w$ be a
 word in $a^{\leq n}(ba^n)^*b \cup \{\epsilon\}$ such
 that $wa^nba^m$ is a good word. 

A word $u$ over $\{a, b\}$ belongs to 
$\LL{wa^nb, wa^nba^m}$ if and only if $u = u_1u_2 u_3 u_4$ with
\begin{enumerate}
\item $u_1 \in \LL{wa^nb, wa^n}$,
\item $u_2 \in \LL{ba^m, a}$,
\item $u_3 \in \LL{ba^m, b}$,
\item $|u_4|_a < m$,
\item $|u_2u_4|_a = 0 {\rm ~mod~} m$,
\item $|u_1|_a (|w|_b + 1) = (|w|_a + n) |u|_b$,
\item ${|u_2|_a +|u_4|_a \over m} - |u_2|_b \leq |u_1| - 
{|u_1|_a (|w| + n) \over (|w|_a + n)}$.
\end{enumerate}
\end{proposition}

\begin{proofWithQed}

\textit{Proof of the ``if part''.} Assume that $u = u_1u_2u_3u_4$ with
$u_1, u_2, u_3, u_4$ verifying Conditions~1 to 7 of the proposition.
Let $\alpha_1, \beta_1, \alpha_2, \beta_2, \alpha_3, \beta_3$ be the
integers (one can verify they are unique) such that:
\begin{itemize}
\item any derivation from $\epsilon$ to $u_1$ by $\qoEtoile{wa^nb, wa^n}$
uses $\alpha_1$ rewriting steps by $\qo{wa^nb}$ and $\beta_1$ steps by
$\qo{wa^n}$;
\item any derivation from $\epsilon$ to $u_2$ by $\qoEtoile{ba^m, a}$
uses $\alpha_2$ rewriting steps by $\qo{ba^m}$  ($\alpha_2 = |u_2|_b$)
and $\beta_2$ steps by $\qo{a}$ ($\beta_2 = |u_2|_a - m |u_2|_b$);
\item any derivation from $\epsilon$ to $u_3$ by $\qoEtoile{ba^m, b}$
uses $\alpha_3$ rewriting steps by $\qo{ba^m}$ ($\alpha_3 = |u_3|_a / m$)
and $\beta_3$ steps by $\qo{b}$.
\end{itemize}

By hypothesis, $|u_2u_4|_a = 0 {\rm ~mod~} m$~: let

\begin{eqnarray}
\label{eq1Gen2}
\beta_2' & = & |u_2u_4|_a /m - |u_2|_b (= (\beta_2 + |u_4|_a) /m).
\end{eqnarray}

Let us observe some relations:
\begin{itemize}
\item We have $|u_1| = \alpha_1 |wa^nb| + \beta_1 |wa^n| = \alpha_1 + 
(\alpha_1 + \beta_1) (|w|+n)$ and\\
$|u_1|_a = \alpha_1 |wa^n|_a + \beta_1 |wa^n|_a = (\alpha_1 + \beta_1) 
(|w|_a+n)$. So
\begin{eqnarray}
\label{eq1Gen1}
\alpha_1 & = & |u_1| - {|u_1|_a (|w|+n) \over |w|_a + n}. 
\end{eqnarray}
\item We also have $|u|_b = (\alpha_1 + \beta_1) |w|_b + \alpha_1 + \alpha_2 + \alpha_3 + \beta_3+ |u_4|_b
= (\alpha_1 + \beta_1) (|w|_b+1) - \beta_1 + \alpha_2 + \alpha_3 + \beta_3+ |u_4|_b$. Since by hypothesis,
$|u_1|_a (|w|_b + 1) = (|w|_a + n) |u|_b$, and since $\alpha_1 + \beta_1 =
{|u_1|_a \over |w|_a + n}$, we have
\begin{eqnarray}
\label{eq3Gen1}
\beta_1 & = & \alpha_2 + \alpha_3 + \beta_3 + |u_4|_b.
\end{eqnarray}
\end{itemize}
We have defined the integers $\alpha_1, \beta_1, \alpha_2, \beta_2', \alpha_3, \beta_3$ in such a
way that:
\begin{itemize}
\item $u_1$ is a shuffle of $\alpha_1$ words $wa^nb$ 
and $\beta_1$ words $wa^n$,
\item $u_2u_4$ is a shuffle of $\alpha_2$ words $ba^m$, $\beta_2'$ words $a^m$ and $|u_4|_b$ words $b$,
\item $u_3$ is a shuffle of $\alpha_3$ words $ba^m$ and $\beta_3$ words $b$.
\end{itemize}
 Since $\beta_1 = \alpha_2 + \alpha_3 + \beta_3 + |u_4|_b$, 
the $\beta_1$ occurrences of
$wa^n$ in $u_1$ can be associated to the $\alpha_2+\alpha_3$ occurrences of
$ba^m$ 
in $u_2u_3$ 
and the $\beta_3 + |u_4|_b$ occurrences of $b$ in $u_3u_4$ 
in order to obtain $\alpha_2+\alpha_3$ occurrences of $wa^nba^m$ and $\beta_3 +
|u_4|_b$ occurrences of $wa^nb$ as subwords of $u$. By Condition~7 and
Relations~(\ref{eq1Gen2}) and (\ref{eq1Gen1}) we have $\beta_2' \leq \alpha_1$.
Thus we can associate $\beta_2'$ occurrences of $wa^nb$ in $u_2$ with the
$\beta_2'$ occurrences of $a^m$ in $u_2u_4$ to construct $\beta_2'$
occurrences of $wa^nba^m$ as subwords in $u$.  So $u$ is the shuffle
of $\beta_2' + \alpha_2 + \alpha_3$ words $wa^nba^m$ and 
$(\alpha_1 - \beta_2') +
\beta_3 + |u_4|_b$ words $wa^nb$ and hence $u \in \LL{wa^nba^m, wa^nb}$.

\medskip

\textit{Proof of the ``only if'' part.}

Assume $u \in \LL{wa^nb, wa^nba^m}$.  Let $\alpha$ and $\beta$ be the
integers (one can verify they are unique) such that any derivation
from $\epsilon$ to $u$ by $\qoEtoile{wa^nb, wa^nba^m}$ uses $\alpha$
rewriting steps by $\qo{wa^nba^m}$ and $\beta$ steps by $\qo{wa^nb}$. An
important remark is that $u(a^m)^\beta \in \LL{wa^nba^m}$.

We have $|u|_a = \alpha|wa^nba^m|_a + \beta|wa^nb|_a = 
(\alpha+\beta)(|w|_a+n) + \alpha m$ and
$|u|_b = (\alpha + \beta) (|w|_b + 1)$. Thus
\begin{eqnarray}
\label{eq4Gen1}
\alpha + \beta &= & {|u|_b \over |w|_b + 1} = {|u|_a \over |w|_a + n} - 
{\alpha m \over
|w|_a + n}.
\end{eqnarray}
 In particular $|u|_b$ is divisible by $|w|_b + 1$, and $|u|_a \geq
 {|u|_b \over |w|_b + 1} (|w|_a + n)$. Let $u_1$ be a prefix of $u$
 such that $|u_1|_a = {|u|_b \over |w|_b + 1} (|w|_a + n) = (\alpha +
 \beta)(|w|_a + n)$. By Lemma~\ref{propertiesNumbering}(2), since
 $u(a^m)^\beta$ belongs to $\LL{wa^nba^m}$, we have $u_1 \in
 \LL{wa^nb,wa^n}$.

Let $s$ be the word such that $u = u_1 s$. By
Lemma~\ref{basicLemmaGeneralized}, $s = u_2u_3u_4$ with $u_2 \in
\LL{ba^m, a}$, $u_3 \in \LL{ba^m, b}$ and $|u_4|_a < m$.

Let us observe that $|u_3|_a = 0 {\rm ~mod~} m$ and
$|s|_a = |u|_a - |u_1|_a = \alpha m = 0 {\rm ~mod~} m$. 
Thus $|u_2u_4|_a = |s|_a - |u_3|_a = 0 {\rm ~mod~} m$.

By Condition~2.2 of Proposition~\ref{carac} applied to 
$n_w = \max(n,m)$ and $f_w = m$, and since $u(a^m)^\beta \in \LL{wa^nba^m}$,
we have 
$|u_3u_4(a^m)^\beta|_a \geq m |u_3u_4(a^m)^\beta|_b = m |u_3u_4|_b$, 
that is, 
$$\beta m + |u_3u_4|_a \geq m|u_3u_4|_b = 
m(|u|_b - |u_1u_2|_b) = m(|u|_b - |u_2|_b - |u_1| + |u_1|_a).$$
The latter inequality  can be
rewritten as $$\beta m +|u|_a - |u_1u_2|_a \geq m \left(
|u|_b - (|u_1|-|u_1|_a)
-|u_2|_b\right),$$ and so 
$$|u_2|_a - m |u_2|_b \leq m|u_1| - \left(m|u|_b + (m+1) |u_1|_a
- (|u|_a + \beta m)\right).$$ 
By recalling that $|u|_b = {|u_1|_a (|w|_b + 1) \over
|w|_a + n}$ and since $$|u|_a + \beta m = (\alpha  + \beta) 
(|w|_a + n + m) = {|u|_b \over
|w|_b +1} (|w|_a + n + m) = {|u_1|_a \over |w|_a +1} (|w|_a + n + 
m),$$ we have
$$\small m|u|_b + (m+1) |u_1|_a - (|u|_a + \beta m) =
 {|u_1|_a \over |w|_a
+n} \left(m(|w|_b +1) + (m+1)(|w|_a+n) - (|w|_a +n+m)\right),$$
which gives 
$$m|u|_b + (m+1) |u_1|_a - (|u|_a + \beta m) =  m {|u_1|_a \over |w|_a +n}
(|w| +n).$$
This shows that $${|u_2|_a \over m} - |u_2|_b \leq |u_1| - {|u_1|_a
\over |w|_a +n} (|w| +n).$$ Now observe that $|u_1| - {|u_1|_a \over |w|_a
+n} (|w| +n) = |u_1| - (\alpha + \beta )(|w|+n)$ is an integer, 
and since $|u_4|_a
< m$ and $|u_2u_4|_a = 0 {~mod~} m$, we have $\lceil {|u_2|_a \over m} \rceil
= {|u_2|_a + |u_4|_a\over m} $. This implies that
$${|u_2|_a + |u_4|_a \over m} - |u_2|_b \leq |u_1| - {|u_1|_a
\over |w|_a +n} (|w| +n).$$
The proof is thus complete
\end{proofWithQed}

\medskip
\noindent
We are now able to prove Proposition \ref{inducStep1}.

\medskip
\begin{proof2}{Proposition \ref{inducStep1}} 
Let $(u_k)_{k \geq 0}$ be a sequence of words in $\LL{wa^nb,
wa^nba^m}$.  By Proposition~\ref{inductiveDecomposition1}, for any $k
\geq 0$, there exist words $u_{1,k}$, $u_{2,k}$, $u_{3,k}$ and
$u_{4,k}$ such that $u_k = u_{1,k} u_{2,k} u_{3,k}u_{4,k}$ with 
\begin{itemize}

\item $u_{1,
k} \in \LL{wa^nb, wa^n}$, 
\item $u_{2,k} \in \LL{ba^m, a}$,
\item $u_{3, k} \in
\LL{ba^m, b}$,
\item $|u_{4, k}|_a < m$, 
\item $|u_{2,k}u_{4,k}|_a = 0 {\rm ~mod~}
m$, 
\item $|u_{1,k}|_a (|w|_b + 1) = (|w|_a + n) |u_k|_b$,
\item ${|u_{2,k}|_a
+|u_{4,k}|_a \over m} - |u_{2,k}|_b \leq |u_{1,k}| - {|u_{1,k}|_a (|w|
+ n) \over (|w|_a + n)}$.
\end{itemize}
Let us define the following integer sequence $(d_k)_{k\geq 0}$: 
for every $k \geq 0$,  $$d_k = |u_{1,k}| - {|u_{1,k}|_a (|w| + n) \over
(|w|_a + n)} - \left({|u_{2,k}|_a +|u_{4,k}|_a \over m} -
|u_{2,k}|_b\right).$$ 

By replacing $(u_k)_{k \geq 0}$ with one of its 
subsequence, we can assume that the sequence $(d_k)_{k \geq 0}$
is non-decreasing.

By hypothesis, $\qoEtoile{wa^nb, wa^n}$ is a wqo on $\LL{wa^nb, wa^n}$,
and by Proposition~\ref{usefulWQO}, $\qoEtoile{ba^m, a}$
(resp. $\qoEtoile{ba^m, b}$) is a wqo on $\LL{ba^m, a}$
(resp. $\LL{ba^m, b}$).  So still replacing $(u_k)_{k \geq 0}$ by a
subsequence, we can assume that, for all $k \geq 0$, 
$$u_{1,k}
\qoEtoile{wa^nb, wa^n} u_{1, k+1},\;\; u_{2,k} \qoEtoile{ba^m, a} u_{2,
k+1},\;\; u_{3,k} \qoEtoile{ba^m, b} u_{3, k+1}.$$  
Moreover, since
$|u_{4, k}|_a$ is bounded,
we can assume that $|u_{4,k}|_a = |u_{4,k+1}|_a$
and since the subsequence ordering is a wqo on $A^*$,  we can assume 
that $u_{4, k}$ is a subword of $u_{4, k+1}$.

The previous arguments imply the existence, for any $k \geq 0$, 
of words $v_{1,k}$, $v_{2,k}$, $v_{3,k}$, $v_{4,k}$ such that 
$$u_{i, k+1} \in u_{i, k} \shuffle v_{i,k}, v_{1, k} \in \LL{wa^nb,
wa^n}, v_{2,k} \in \LL{ba^m, a}, v_{3, k} \in \LL{ba^m, b}, |v_{4,
k}|_a =0.$$ The equality $|v_{2,k}v_{4,k}|_a = 0 {\rm ~mod~} m$ easily
follows from $|u_{2,k}u_{4,k}|_a = 0 {\rm ~mod~} m$ and
$|u_{2,k+1}u_{4,k+1}|_a = 0 {\rm ~mod~} m$.  We have $|v_{1,k}|_a =
|u_{1, k+1}|_a - |u_{1,k}|_a$ and, taking $v_k =
v_{1,k}v_{2,k}v_{3,k}v_{4,k}$, $|v_k|_b = |u_{k+1}|_b - |u_k|_b$.
Since $|u_{1,j}|_a (|w|_b + 1) = (|w|_a + n) |u_j|_b$ for $j \in \{k,
k+1\}$, we can deduce that $|v_{1,k}|_a (|w|_b + 1) = (|w|_a + n)
|v_k|_b$.  By the fact that the sequence $(d_k)_{k\geq 0}$ is
non-decreasing, we have
$${|v_{2,k}|_a +|v_{4,k}|_a \over m} - |v_{2,k}|_b \leq |v_{1,k}| -
{|v_{1,k}|_a (|w| + n) \over (|w|_a + n)}.$$
Now,  by applying Proposition~\ref{inductiveDecomposition1} to the words
$v_{k}$, we 
have $v_k \in
\LL{wa^nb, wa^nba^m}$. Since,  for all
$k \geq 0$, $u_{k+1} \in u_k \shuffle v_k$, the latter condition gives
$u_{k}\;\qoEtoile{wa^nb, wa^nba^m}\; u_{k+1}$. Therefore
$\qoEtoile{wa^nb, wa^nba^m}$is a wqo on $\LL{wa^nb,
wa^nba^m}$.
\end{proof2}

\subsection{A second inductive result}

The aim of this section is to prove the next result which proof is
based on the characterization provided by
Proposition~\ref{inductiveDecomposition2}. 

\begin{proposition}
\label{inducStep2}
Let $n \geq 1$ be an integer and let $w$ be
 a word in $a^{\leq n}(ba^n)^*$.
If $\qoEtoile{wb, wba^n}$ is a wqo on
$\LL{wb, wba^n}$ then $\qoEtoile{wba^n, wba^nb}$ is a wqo
on $\LL{wba^n, wba^nb}$.
\end{proposition}

\begin{proposition}
 \label{inductiveDecomposition2}
Let $n \geq 1$ be an integer and let 
$w \in a^{\leq n}(ba^n)^*$.  A word $u$ belongs to $\LL{wba^n,
wba^nb}$ if and only if $u = u_1 u_2 u_3 u_4 u_5 u_6$ with\footnote{the value of $\overline{\delta}_{u_2u_3, \epsilon}$ is 0 if $u_2u_3 =
\epsilon$ and 1 otherwise}:

\begin{enumerate}
\item $u_1b^{|u_2|_b} \in \LL{wb, wba^n}$,
\item $|u_1u_2|_b (|w|_a + n) = |u|_a (|w|_b+1)$, 
\item $u_2u_3 = \epsilon$ or $|u_2u_3|_a = n$,
\item $|u_4|_a < n$,
\item $u_5 \in \LL{a^nb,b}$,
\item $u_6 \in \LL{a^nb,a}$,
\item\label{item8} $|u_3|_b \leq 
{1 \over n}\left[|u_1|_a - {|u_1u_2|_b \over |w|_b + 1} |w|_a\right]$,
\item\label{item9} $|u_5|_b - {|u_5|_a \over n} + |u_3u_4|_b \leq {1 \over n}
\left[|u_1|_a - {|u_1u_2|_b \over |w|_b + 1} |w|_a\right] +
\overline{\delta}_{u_2u_3, \epsilon}$,
\item\label{item10} ${|u|_a - |u_1|_a \over n} \geq |u_2|_b + \overline{\delta}_{u_2u_3, \epsilon}$.
\end{enumerate}
\end{proposition}

\begin{proofWithQed} 

 \textit{Proof of the ``if'' part.} Assume first that $u$ can be 
 factorized in the product of six words
 satisfying the properties of the proposition.
Let $\alpha_1$, $\beta_1$, $\alpha_5$, $\beta_5$, $\alpha_6$, $\beta_6$
be the integers (one can verify they are
unique) such that:
\begin{itemize}
\item any derivation from $\epsilon$ to $u_1b^{|u_2|_b}$ by 
$\qoEtoile{wba^n, wb}$
uses $\alpha_1$ rewriting steps by $\qo{wba^n}$ and $\beta_1$ steps by
$\qo{wb}$;
\item any derivation from $\epsilon$ to $u_5$ by $\qoEtoile{a^nb, b}$
uses $\alpha_5$ rewriting steps by $\qo{a^nb}$ ($\alpha_5 = |u_5|_a / n$)
and $\beta_5$ steps by $\qo{b}$ ($\beta_5 = |u_5|_b - \alpha_5$);
\item any derivation from $\epsilon$ to $u_6$ by $\qoEtoile{a^nb, a}$
uses $\alpha_6$ rewriting steps by $\qo{a^nb}$ ($\alpha_6 = |u_6|_b$)
and $\beta_6$ steps by $\qo{a}$ ($\beta_6 = |u_6|_a - n \alpha_6$).
\end{itemize}

Let us observe some relations:
\begin{itemize}
\item We have $|u_1|_a = 
\alpha_1 |wba^n|_a + \beta_1 |wb|_a = n \alpha_1 + (\alpha_1 + \beta_1)
|w|_a$ and $|u_1u_2|_b = (\alpha_1+\beta_1) (|w|_b + 1)$. So we have 
\begin{eqnarray}
\label{eq1Gen1bis}
\alpha_1 & = & {1 \over n}
\left[|u_1|_a - {|u_1u_2|_b \over |w|_b + 1} |w|_a\right].
\end{eqnarray}
Thus Properties~\ref{item8} and \ref{item9} can be rephrased 
$|u_3|_b \leq \alpha_1$ and 
$\beta_5 + |u_3u_4|_b \leq \alpha_1 + 
\overline{\delta}_{u_2u_3, \epsilon}$ respectively.
\item We also have $|u|_a = 
\alpha_1 (|w|_a+n) + \beta_1 |w|_a + |u_2u_3u_4u_6|_a + n \alpha_5
= (\alpha_1 + \beta_1 ) (|w|_a+n) - \beta_1 n + |u_2u_3u_4u_6|_a + n \alpha_5$.
Thus from Property~2 and the equality 
$|u_1u_2|_b = (\alpha_1+\beta_1) (|w|_b + 1)$, we have:
\begin{eqnarray}
\label{eq3Gen1bis}
\beta_1 n & = & |u_2u_3u_4u_6|_a + n \alpha_5.
\end{eqnarray}
\end{itemize}

We first consider the case where $u_2u_3 = \epsilon$.
The previous equality shows that $|u_4u_6|_a$ is a multiple of
$n$. Moreover the $\beta_1$ occurrences of $wb$ in $u_1$ can be associated
to the $\alpha_5 + \alpha_6$ occurrences of $a^nb$ in $u_5u_6$ and 
to the $|u_4u_6|_a /n -  \alpha_6$ remaining occurrences of $a$ in 
$u_4u_6$ to form $\alpha_5 + \alpha_6$ occurrences of
$wba^nb$ and $(|u_4u_6|_a - n \alpha_6) / n$ occurrences of
$wba^n$.
We have seen
as a consequence of Relation~(\ref{eq1Gen1bis}), that $\beta_5 +
|u_4|_b \leq \alpha_1$. 
Thus $\beta_5 + |u_4|_b$ occurrences 
of $wba^n$ in
$u_1$ can be associated to some corresponding $b$ in
$u_4u_5$ to form some occurrences of $wba^nb$ in $u$.
Finally we have shown that $u$ is the shuffle
of $\alpha_5 + \alpha_6 + \beta_5 + |u_4|_b$ of $wba^nb$ and
$(|u_4u_6|_a - n \alpha_6) / n + \alpha_1 - (\beta_5 + |u_4|_b)$ occurrences
of $wba^n$.

\medskip

We now consider the case where $u_2u_3 \neq \epsilon$.  We start
exploiting Property~\ref{item10}~: ${|u|_a - |u_1|_a \over n} \geq |u_2|_b + 1$.
We already know that $|u_1u_2|_b = (\alpha_1 + \beta_1) (|w|_b + 1)$,
so by Property~2, $|u|_a = (\alpha_1 + \beta_1) (|w|_a + n)$. Moreover
$|u_1|_a = (\alpha_1 + \beta_1) |w|_a + \alpha_1 n = |u|_a - \beta_1 n$.  
Thus Property~\ref{item10} can be rewritten
$\beta_1 \geq |u_2|_b + 1$.
This means that at least one occurrence of the $\beta_1$ occurrences
of $wb$ in $u_1b^{|u_2|_b}$ is completely included as a subword in $u_1$.
There exists a subword $x_1$ of $u_1$ such that
$x_1b^{|u_2|_b} \in \LL{wb, wba^nb}$, $|x_1|_b = |u_1|_b - |wb|_b$,
$|x_1|_a = |u_1|_a - |w|_a$. Let $u_1' = x_1b^{|u_2|_b}$,
$u_2' = u_3' = \epsilon$.

If $|u_4|_b \neq 0$, let $x_4$ be a subword of $u_4$ with $|x_4|_a =
|u_4|_a$, $|x_4|_b = |u_4|_b-1$ and let $u_4' = b^{|u_3|_b}x_4$, $u_5'
= u_5$, $u_6' = u_6$.  If $|u_4|_b = 0$, let $u_4' = b^{|u_3|_b}u_4$.
If $|u_4|_b = 0$ and $|u_5|_b - {|u_5|_a \over n} \neq 0$, let $u_5'$
be the subword of $u_5$ obtained by erasing the first occurrence of
$b$ in $u_5$ and let $u_6' = u_6$.
If $|u_4|_b = 0$ and $|u_5|_b - {|u_5|_a \over n} = 0$,
let $u_5' = u_5$, $u_6' = u_6$.
Finally let $u' = u_1'u_2'u_3'u_4'u_5'u_6'$. 

By the previous construction, the word $u$ is the shuffle of $u'$ and
one of the two words $wba^n$ or $wba^nb$ (constituted with a subword
$wb$ in $u_1$, the $|u_2u_3|_a = n$ occurrences of $a$ in $u_2u_3$, and
possibly a $b$ occurring in $u_4u_5$).  We now verify that the words
$u'$, $u_1'$, $u_2'$, $u_3'$, $u_4'$, $u_5'$, $u_6'$ satisfy
Properties 1 to \ref{item10} of the Proposition.  We have already said
that $u_1'b^{|u_2'|_b} = u_1' \in \LL{wb, wba^n}$.  We have
$|u_1'u_2'|_b = |u_1u_2|_b - (|w|_b+1)$ and $|u'|_a = |u|_a -
(|w|_a+n)$ which gives  $|u_1'u_2'|_b(|w|_a+n) = |u'|_a(|w|_b+1)$.  The
verification (left to the reader) of Properties 3 to \ref{item8} and
\ref{item10} are immediate.

Let us prove Property 8.

Let $X = |u_5|_b - {|u_5|_a \over n} + |u_3u_4|_b$, $Y = {1 \over
n}\left[|u_1|_a - {|u_1u_2|_b \over |w|_b + 1} |w|_a\right]$, $X' =
|u_5'|_b - {|u_5'|_a \over n} + |u_3'u_4'|_b$, $Y' = {1 \over n}
\left[|u_1'|_a - {|u_1'u_2'|_b \over |w|_b + 1} |w|_a\right]$.  By
Property \ref{item10} for $u$, we have $X \leq Y +1$ and we want to
prove that $X' \leq Y'$.
As a consequence of the definition of the words $u'_{i}$, it is
easily seen that $$X = X' + 1\;\mbox{or}\;X = X'.$$
Moreover, one can easily verify that the last equality occur only if
$$ |u_4|_b =  |u_5|_b - {|u_5|_a \over n} = 0,$$
which gives
$$X = |u_3|_b.$$
On the other hand, since 
$|u_1|_a = |u_1'|_a + |w|_a$ and $|u_1u_2|_b = |u_1'u_2'|_b + (|w|_b +
1)$, we have $$Y = Y'.$$ 
By the latter equality, $X = X' + 1$ immediately gives $X' \leq Y'$,
while, if $X = X'$, by Property~7, $X \leq Y$, that is $X' \leq
Y'$.

Thus the words $u'$, $u_1'$, $u_2'$, $u_3'$, $u_4'$, $u_5'$, $u_6'$
satisfy Properties 1 to \ref{item10} of the Proposition with $u_2'u_3' = \epsilon$.
By the previous case, $u' \in \LL{wba^n, wba^nb}$ and so 
$u \in \LL{wba^n, wba^nb}$.

\medskip

 \textit{Proof of the `` only if'' part.}
Let us first note that, by definition of $w$, there exists an
integer $i_w$ between $0$ and $n$ such that $wba^nb =
a^{i_w}b(a^nb)^{|w|_b+1}$.

Assume $u$ belongs to $\LL{wba^n,wba^nb}$.
There exist unique integers $\alpha$ and $\beta$ such that any
derivation from $\epsilon$ to $u$ by $\qoEtoile{wba^n,wba^nb}$ uses
$\alpha$ derivation steps by $\qo{wba^nb}$ and $\beta$ derivation
steps by $\qo{wba^n}$. We have:
$$|u|_a = (\alpha + \beta) (|w|_a + n), {\rm ~and~}$$
$$|u|_b = (\alpha + \beta) (|w|_b + 1) + \alpha.$$
In particular, $|u|_a$ is divisible by $|w|_a + n$ and
$|u|_b \geq {|u|_a (|w|_b + 1) \over |w|_a + n}$.

Let $p$ be a prefix of $w$ such that $|p|_b = {|u|_a (|w|_b + 1) \over
  |w|_a + n} (= (\alpha + \beta) (|w|_b + 1))$, and let $s$ be the
word such that $u = ps$. 
Since $i_w \leq n$, the
$(\alpha + \beta)^{\rm th}$ occurrence of the letter $b$ is preceded
by at least $(\alpha + \beta)i_w$ occurrences of the letter $a$.  Let
$u_1$ be the longest prefix of $p$ such that $|u_1|_a \geq (\alpha +
\beta)i_w$ and $|u_1|_a - (\alpha + \beta)i_w \mod n = 0$, and let $u_2$
be the word such that $p = u_1u_2$: by construction $u_2 =
\epsilon$, or, $u_2$ begins with the letter $a$ and $0 < |u_2|_a <
n$.
Observe  $|u|_a - (\alpha + \beta)i_w = 0 \mod n$. So we can consider the
shortest prefix $u_3$ of $s$ such that $|u_2u_3|_a = 0 \mod n$. We observe
that if $u_2 = \epsilon$ then $u_3 = \epsilon$, and otherwise
$u_3 \neq \epsilon$ and $|u_2u_3|_a = n$.

By
Lemma~\ref{basicLemmaGeneralized}, there exist words $u_4$, $u_5$,
$u_6$ such that $\tilde{s} = \tilde{u}_6\tilde{u}_5\tilde{u}_4$ with
$\tilde{u}_6 \in \LL{ba^n, a}$, $\tilde{u}_5 
\in \LL{ba^n, b}$ and $|\tilde{u}_4|_a < n$. Thus
$s = u_4 u_5 u_6$, $|u_4|_a < n$, $u_5 \in \LL{a^nb, b}$, $u_6 \in
\LL{a^nb, a}$.

Up to now, we have constructed words $u_1$, \ldots, $u_6$ verifying
required Properties 2 to 6.  We
have $|u_1|_a \mod n = |u|_a \mod n = (\alpha + \beta)i_w \mod n$,
$|u_2u_3|_a = 0 \mod n$ and $|u_5|_a = 0 \mod n$: thus $|u_4u_6|_a = 0 \mod n$.
We now concentrate our efforts on Properties~1 and \ref{item8} to
\ref{item10}.  The word $ub^\beta$ belongs to $\LL{wba^nb}$ and
$|ub^\beta| = (\alpha + \beta) |wba^nb|$.  Let us recall
that $wba^nb = a^{i_w}b(a^nb)^{|w|_b+1}$. Condition 2.1
of Proposition~\ref{carac} shows that, taking $x = \alpha + \beta =
{|ub^\beta| \over |wba^nb|}$, $|p|_a \geq i_w x + n (|p|_b - x)$.
But $|p|_a = |ub^\beta|_a - |s|_a = x |wba^nb|_a - |s|_a = x(i_w +
(|w|_b + 1)n) - |s|_a = x i_w + n |p|_b - |s|_a$.  Thus ${|s|_a \over
n} \leq x$.

By Proposition~\ref{carac} and Lemma~\ref{propertiesNumbering}, we
know that $ub^\beta$ is the shuffle of the $(\alpha+\beta)$ words
$(ub^\beta)(i)$ ($1 \leq i \leq \alpha + \beta$) defined just before
Lemma~\ref{propertiesNumbering}.  Let us recall that $(ub^\beta)(i)$
is the subword of $ub^\beta$ constituted by the letters in position in
$P(i)$. Let $p(i)$ be the subword of $p$ constituted by the letters in
position in $P(i) \cap \{1, \ldots, |p|\}$, and let $s(i)$ be the
words such that $(ub^\beta)(i) = p(i) s(i)$.

The proof is divided into the following two cases according
to the  value of $|s|_a \mod n = |u_3|_a$.

\medskip

{\em Case~ $|s|_a = 0 \mod n $}. In particular $u_2 = u_3 = \epsilon$.
In this case, Properties ~\ref{item8} and \ref{item10} are trivially
satisfied.

Let $y = {|s|_a \over n}$.  By the construction of the
$(ub^\beta)(i)$'s (and in particular of the values of elements of $P(i)$) 
we have that:
\begin{itemize}
\item $p(i) = wba^n$, $s(i) = b$, for $1 \leq i \leq x-y$,
\item $p(i) = wb$, $s(i) = a^nb$, for $x-y+1 \leq i \leq x$.
\end{itemize}
This implies  $p = u_1b^{|u_2|_b} \in \LL{wba^n, wb}$
and $sb^\beta \in \LL{a^nb,b}$. In particular we have
Property~1.

There exist unique integers $\alpha_5$
and $\beta_5$ such that any derivation from $\epsilon$ to $u_5$ by
$\qoEtoile{a^nb,b}$ uses $\alpha_5$ derivation steps by $\qo{a^nb}$ and
$\beta_5$ derivation steps by $\qo{b}$, and there exist unique integers
$\alpha_6$ and $\beta_6$ such that any derivation from $\epsilon$
to $u_6$ by $\qoEtoile{a^nb,a}$ uses $\alpha_6$ derivation steps by
$\qo{a^nb}$ and $\beta_6$ derivation steps by $\qo{a}$. In particular,
we have
$\beta_5 = |u_5|_b - {|u_5|_a \over n}.$

Let us prove that
$\beta_5 + |u_4|_b \leq x-y$.  By Lemma~\ref{basicLemmaGeneralized},
the value of $\alpha_5 + \alpha_6$ is the greatest number $z$ such
that $u_4u_5u_6$ can be viewed as the shuffle of $z$ occurrences of
$a^nb$ with some occurrences of $a$ and some occurrences of $b$.  Due
to the fact that $sb^\beta = u_4u_5u_6b^\beta$ is the shuffle of $y$
occurrences of $a^nb$ and $(x-y)$ occurrences of $b$, we get $y \leq
\alpha_5 + \alpha_6 + \beta$.  It follows: $x = |sb^\beta|_b = 
|u_4u_5u_6b^\beta|_b = |u_4|_b + \alpha_5 + \beta_5 + \alpha_6 + \beta
\geq |u_4|_b +
\beta_5 + y$. So $x-y \geq \beta_5 + |u_4|_b$.
 
Since $p = u_1$, $p$ is the shuffle of $x-y$ occurrences of $wba^n$ and $y$
occurrences of $wb$. We have $|p|_a = (x-y)(|w|_a + n) + y |w|_a = x
|w|_a + n (x-y)$ and $|p|_b = (x-y) |wb|_b + y |wb|_b = x(|w|_b + 1)$.
Thus $n(x-y) = |p|_a - {|p|_b |w|_a \over |w|_b + 1}$.
Since $u_2 = u_3 = \epsilon$, $p = u_{1}u_{2}$,
¥$\beta_5 + |u_4|_b \leq x-y$ and $\beta_5 = 
|u_5|_b - {|u_5|_a \over n}$, we have
$$|u_5|_b - {|u_5|_a \over n} + |u_3u_4|_b \leq {1 \over n}
\left[|u_1|_a - {|u_1u_2|_b \over |w|_b + 1} |w|_a\right].$$
Hence Property~\ref{item9} is proved.

\medskip

{\em Case~ $|s|_a \neq 0 \mod n $.}  We still have $\alpha + \beta = x \geq
{|s|_a \over n}$.  Let $y = \lfloor {|s|_a \over n} \rfloor$: $0 \leq
y < x$. By construction of the $(ub^\beta)(i)$'s,
\begin{itemize}
\item $p(i) = wba^n$, $s(i) = b$, for $1 \leq i \leq x-y-1$;
\item $p(x-y) = wba^r$, $s(x-y) = a^{n-r}b$ for an integer $r$, 
$1 \leq r < n$;
\item $p(i) = wb$, $s(i) = a^nb$ for $x-y+1 \leq i \leq x$.
\end{itemize}
It follows that $|u_2|_a = r$ and $u_1b^{|u_2|_b} \in \LL{wba^n, wb}$.
Hence we have proved Property~1.

Let us recall that $s = u_3u_4u_5u_6$ and $sb^\beta$ is the shuffle of
the $x$ words $s(i)$. Since $b^{|u_3|_b}u_4u_5u_6b^\beta$ is the
shuffle of $y$ occurrences of $a^nb$ and $(x-y)$ occurrences of $b$,
by using an argument similar to that of the previous case, we have that $|u_5|_b -
{|u_5|_a \over n} + |u_3u_4|_b \leq x-y$.

Here $p$ is the shuffle of $x-y-1$ occurrences of $wba^n$, one
occurrence of $wba^r$ and $y$ occurrences of $wb$.  Thus $|u_1u_2|_a =
|p|_a = (x-y-1)(|w|_a+n) + (|w|_a+r) + y |w|_a$ with $r = |u_2|_a$.
So $|u_1|_a = x|w|_a + (x-y)n - n$.
Since $x = |u_1u_2|_b / (|w|_b + 1)$, we get
$n(x-y) = |u_1|_a - {|u_1u_2|_b  |w|_a \over |w|_b + 1} + n$.
And so, we have Property~\ref{item9}:
$$|u_5|_b - {|u_5|_a \over n} + |u_3u_4|_b \leq {1 \over n}
\left[|u_1|_a - {|u_1u_2|_b \over |w|_b + 1} |w|_a\right] + 1.$$

By construction of the words $s(i)$'s, 
for all $i$ such that $x-y+1 \leq i \leq x$, 
the occurrences of the letter $a$ in $s(i)$ 
appear in $ub^\beta$ after the occurrences of the letter $a$ in
$s(x-y)$. More precisely, for an integer $i \geq x-y+1$,
if the letter $a$ occurs in 
$ub^\beta$ at two positions $j$ and $k$ with $j \in P(x-y) \cap
\{|p|+1, \ldots,|u|\}$, and $k \in P(i) \cap \{|p|+1, \ldots,|u|\}$,
then $j < k$. On the other hand, by definition of $u_3$, the last
letter of $u_3$ is $a$. Hence for any $i \geq x-y$, each letter $b$
in $s(i)$ cannot occur in $u_3$, so that $|u_3|_b < x-y$. 
Therefore, we have $$|u_3|_b < {1 \over n}
\left[|u_1|_a - {|u_1u_2|_b \over |w|_b + 1} |w|_a\right] + 1,$$
and  Property~\ref{item8} is proved.
 
By construction, $u_2$ starts with the letter $a$. It follows that
$u_1$ contains all the $b$'s occurring in the $p(i)$'s for $1 \leq i
\leq x-y$, and those occurring in the prefix $w$ of the $p(i)$'s for
$x-y+1 \leq i \leq x$, that is, $|u_1|_b \geq (x-y) |wb|_b + y |w|_b =
x |wb|_b - y = |u_1u_2|_b - y$ and, hence, $y \geq |u_2|_b$.  But $|u_1|_a =
x|w|_a + (x-y-1)n = x(|w|_a+n) - (y+1) n = |u|_a -
(y+1)n$. Consequently, we have Property~\ref{item10}:
$${|u|_a - |u_1|_a \over n} \geq |u_2|_b + 1 = |u_2|_b +
\overline{\delta}_{u_2u_3, \epsilon}.$$
\end{proofWithQed}
 
\begin{proof2}{Proposition~\ref{inducStep2}}
The proof follows the same scheme of that of Proposition~\ref{inducStep1}
but the arguments used here are  more technical.

Let $(u_k)_{k \geq 0}$ be a sequence of words in $\LL{wba^n,
wba^nb}$.  By Proposition~\ref{inductiveDecomposition2}, for any $k
\geq 0$, there exist six words $u_{1,k}$, \ldots, $u_{6,k}$
such that $u_k = u_{1,k} \ldots u_{6,k}$ with
\begin{itemize}
\item $u_{1,k}b^{|u_{2,k}|_b} \in \LL{wb, wba^n}$,
\item
$|u_{1,k}u_{2,k}|_b (|w|_a + n) = |u_k|_a (|w|_b+1)$, 
\item
$u_{2,k}u_{3,k} = \epsilon$ or 
$|u_{2,k}u_{3,k}|_a = n$,
\item
$|u_{4,k}|_a < n$,
\item
$u_{5,k} \in \LL{a^nb,b}$,
\item
$u_{6,k} \in \LL{a^nb,a}$,
\item
$|u_{3,k}|_b \leq 
{1 \over n}\left[|u_{1,k}|_a - {|u_{1,k}u_{2,k}|_b \over 
|w|_b + 1} |w|_a\right]$,
\item
$|u_{5,k}|_b - {|u_{5,k}|_a \over n} + |u_{3,k}u_{4,k}|_b \leq {1 \over n}
\left[|u_{1,k}|_a - {|u_{1,k}u_{2,k}|_b \over |w|_b + 1} |w|_a\right] +
\overline{\delta}_{u_{2,k}u_{3,k}, \epsilon}$,
\item ${|u_{k}¥|_a - |u_{1,k}|_a \over n} \geq |u_{2,k}|_b +
 \overline{\delta}_{u_{2,k}u_{3,k}, \epsilon}$.
\end{itemize}
Now let us define the following three sequences of integers: for 
every $k\geq 0$,
$$d_{1,k} = {1 \over n}\left[|u_{1,k}|_a - {|u_{1,k}u_{2,k}|_b \over 
|w|_b + 1} |w|_a\right] - |u_{3,k}|_b,$$
$$d_{2, k} = {1 \over n}
\left[|u_{1,k}|_a - {|u_{1,k}u_{2,k}|_b \over |w|_b + 1} |w|_a\right] +
\overline{\delta}_{u_{2,k}u_{3,k}, \epsilon} - \left(|u_{5,k}|_b - {|u_{5,k}|_a \over n} 
+ |u_{3,k}u_{4,k}|_b\right),$$
$$d_{3,k} = {|u|_a - |u_{1,k}|_a \over n} - \left(|u_{2,k}|_b +
 \overline{\delta}_{u_{2,k}u_{3,k}, \epsilon}\right).$$


By hypothesis, $\qoEtoile{wba^n, wb}$ is a wqo on $\LL{wba^n, wb}$,
and by Proposition~\ref{usefulWQO}, $\qoEtoile{a^nb, b}$
(resp. $\qoEtoile{a^nb, a}$) is a wqo on $\LL{a^nb, b}$
(resp. $\LL{a^nb, a}$).

By the fact that the subsequence ordering is a wqo on $A^*$¥ and 
by taking a suitable subsequence of $(u_k)_{k \geq 0}$, 
we can assume that, for all
$k \geq 0$, the following conditions are satisfied: 
\begin{itemize}
\item
$u_{1,k} \qoEtoile{wba^n, wb} u_{1, k+1}$,
\item $u_{i, k}$
is a subword of $u_{i, k+1}$, for $i = 2, 3, 4$,
\item $|u_{i,k}|_a = |u_{i,k+1}|_a$, for $i = 2, 3, 4$, 
\item $u_{5,k} \qoEtoile{a^nb, b} u_{5,k+1}$, 
\item $u_{6,k} \qoEtoile{a^nb, a} u_{6, k+1}$,
\item $d_{i, k} $ is non-decreasing for $i = 1, 2, 3$. 
\end{itemize}
We have $|u_{2,k}u_{3,k}|_a =
|u_{2,k+1}u_{3,k+1}|_a$ and so $\overline{\delta}_{u_{2,k}u_{3,k}, \epsilon}
= \overline{\delta}_{u_{2,k+1}u_{3,k+1}, \epsilon}$.

From the previous conditions, for any $k \geq 0$, we can easily
deduce the existence of
words $v_{1,k}, v_{2,k}, v_{3,k}, v_{4,k}, v_{5,k}, v_{6,k}$, such that

$$u_{i, k+1} \in u_{i, k} \shuffle v_{i,k}, \;\;\;\; 
v_{1, k}b^{|v_{2, k}|_b} \in \LL{wa^nb, wa^n},\;\;\;\;
|v_{i, k}|_a = 0,$$ {for} $i = 2, 3, 4$ and 
$$v_{5,k} \in \LL{a^nb, b}, \;\;\;\;
v_{6, k} \in \LL{a^n, a}.$$
Let $v_{1,k}' = v_{1, k}b^{|v_{2, k}|_b}$,
$v_{2,k}' = \epsilon$,
$v_{3,k}' = \epsilon$,
$v_{4,k}' = b^{|v_{3, k}|_b}v_{4,k}$,
$v_{5,k}' = v_{5,k}$
and $v_{6,k}' = v_{6,k}$.

By using an argument similar to that 
of the proof of Proposition~\ref{inducStep1}, we can deduce that,
for all $k\geq 0$, 
the words $v_k = v_{1,k}\ldots v_{6,k} = v_{1,k}'\ldots v_{6,k}'$ satisfy all the properties 
of Proposition~\ref{inductiveDecomposition2}, and therefore $v_k' \in
\LL{wba^{n}¥, wba^{n}b}$. This implies that, for all $k\geq 0$,
$v_k \in \LL{wba^{n}¥, wba^{n}b}$.
Since,  for all $k\geq 0$, $u_{k+1} \in u_k \shuffle v_k$,
the latter implies that $u_{k}¥\qoEtoile{wba^n, wba^nb} u_{k+1}$, that 
is $\qoEtoile{wba^n, wba^nb}$ is a wqo on $\LL{wba^n,
wba^nb}$.
\end{proof2}

\subsection{Proof of the ``if'' part of Theorem~\ref{mainTheorem}}

From the results of the previous section we can deduce:
\begin{theorem}
\label{parentOfMainResult}
For any integers $n, m \geq 1$, and for any word $w$ in $a^{\leq
n}(ba^n)^*b \cup \{\epsilon\}$ such that $wa^nba^m$ is a good word,
one has:
\begin{enumerate}
\item $\qoEtoile{wa^n,wa^nb}$ is a wqo on $\LL{wa^n, wa^nb}$;
\item $\qoEtoile{wa^nb, wa^nba^m}$ is a wqo on $\LL{wa^nb, wa^nba^m}$.
\end{enumerate}
\end{theorem}

\begin{proofWithQed}
We act by induction on $|w|_b$.

When $|w|_b = 0$, $w = \epsilon$ and we know by Proposition~\ref{usefulWQO}
that $\qoEtoile{a^n,a^nb}$ is a wqo on $\LL{a^n, a^nb}$.
By Proposition~\ref{inducStep1}, we deduce that
$\qoEtoile{a^nb, a^nba^m}$ is a wqo on $\LL{a^nb, a^nba^m}$.

Assume now $|w|_b \geq 1$. Then
$w = a^hb$ with $0 \leq h \leq n$ or $w =
w'a^nb$ with $w' \in a^{\leq n}(ba^n)^*b$.  If $w = b$, then by
Proposition~\ref{usefulWQO}, $\qoEtoile{b, ba^n}$ is a wqo on $\LL{b,
ba^n}$.  In the other cases, by inductive hypothesis, $\qoEtoile{w,
wa^n}$ is a wqo on $\LL{w, wa^n}$.  So in all cases by
Proposition~\ref{inducStep2}, $\qoEtoile{wa^n, wa^nb}$ is a wqo on
$\LL{wa^n, wa^nb}$, and by Proposition~\ref{inducStep1}, we
deduce that $\qoEtoile{wa^nb, wa^nba^m}$ is a wqo on
$\LL{wa^n, wa^nba^m}$.
\end{proofWithQed}
\begin{corollary}
\label{corMainResult} Let $n \geq 1$ be an integer.
For any word $w$ in
$a^{\leq n}(ba^n)^*ba^{\leq n}$,
$\qoEtoile{w}$ is a wqo on $\LL{w}$.
\end{corollary}

\begin{proofWithQed}
The result is immediate if $|w|_b = 0$.  Assume from now on $|w|_b >
0$.

First we consider the case where $w$ ends with $b$. Two cases are
possible: $w = a^mb$ with $1 \leq m \leq n$ or $w = w'ba^nb$ with $w'$
in $a^{\leq n}(ba^n)^*$.  If $w = a^mb$, the result is stated in
Proposition~\ref{propFS05}.\\ Assume $w = w'ba^nb$. By
Theorem~\ref{parentOfMainResult}, we know that $\qoEtoile{w'ba^n,
w'ba^nb}$ is a wqo on $\LL{w'ba^n, w'ba^nb}$.  Let $(u_k)_{k \geq 0}$
be a sequence of words in $\LL{w'ba^nb}$. Since $\LL{w'ba^nb}
\subseteq \LL{w'ba^n, w'ba^nb}$, $u_k \in \LL{w'ba^n, w'ba^nb}$ and so
we can replace the sequence $(u_k)_{k \geq 0}$ by a subsequence such
that $u_k \qoEtoile{w'ba^n, w'ba^nb} u_{k+1}$ for each $k \geq 0$. For
any $k$ this means there exists a word $v_k$ in $\LL{w'ba^n,
w'ba^nb}$ such that $u_{k+1} \in u_k \shuffle v_k$.  The word $v_k$ is
the shuffle of $\alpha_k$ occurrences of $w'ba^n$ and $\beta_k$
occurrences of $w'ba^nb$, and the words $u_k$ and $u_{k+1}$ are the
shuffle of $\gamma_k$ and $\gamma_{k+1}$  occurrences of
$w'ba^nb$ respectively. From $|v_k|_a  = |u_{k+1}|_a - |u_k|_a$ and
$|v_k|_b  = |u_{k+1}|_b - |u_k|_b$, we deduce respectively
$\alpha_k + \beta_k = \gamma_{k+1} - \gamma_k$ and
$(\gamma_{k+1} - \gamma_k) |w'ba^nb|_b = (\alpha_k + \beta_k) |w'ba^nb|_b -
\alpha_k$ which imply $\alpha_k = 0$, that is, $v_k \in \LL{w'ba^nb}$.
Hence $u_k \qoEtoile{w'ba^nb} u_{k+1}$, so that  $\qoEtoile{w'ba^nb}$ is a wqo
on $\LL{w'ba^nb}$.

Now we consider the case where $w$ ends with $a$~ so that  $w = w'ba^m$ with
$w' \in a^{\leq n}(ba^n)^* \cup \{\epsilon\}$ and $n \geq m \geq
1$. By Theorem~\ref{parentOfMainResult}(2),
$\qoEtoile{w'b,w'ba^m}$ is a wqo on
$\LL{w'b,w'ba^m}$. The proof ends as in the previous case.
\end{proofWithQed}

\medskip

We are now able to prove the ``if''  part of Theorem~\ref{mainTheorem}.
\begin{proof2}{the ``if'' part of Theorem~\ref{mainTheorem}}
Assume $w$ is a word such that $w$, $\tilde{w}$, $E(w)$ and
$E(\tilde{w})$ have no factor of the two possible forms~1 and 2 of
Definition \ref{badw}.  By Lemma~\ref{goodWords}, we know that $$w \in
\{\epsilon\} \cup \bigcup_{n \geq 0} a^{\leq n}(ba^n)^*ba^{\leq n}
\cup \bigcup_{n \geq 0} b^{\leq n}(ab^n)^*ab^{\leq n}.$$ The result is
trivial if $|w|_a = 0$ or $|w|_b = 0$ and stated by
Corollary~\ref{corMainResult} if $w \in a^{\leq n}(ba^n)^*ba^{\leq n}$
with $n \geq 1$. The case $w \in b^{\leq n}(ab^n)^*ab^{\leq n}$ with
$n \geq 1$ is treated as the previous case by exchanging the role of
$a$ and $b$.
\end{proof2}

 \end{document}